\documentclass[12pt]{article}
\usepackage{amsfonts}
\usepackage[centertags]{amsmath}
\usepackage{amssymb}
 \usepackage{bookmark}
\def\ba{\begin{array}}
\def\ea{\end{array}}
\def\be{\begin{equation}}
\def\ee{\end{equation}}
\def\bea{\begin{eqnarray}}
\def\eea{\end{eqnarray}}
 \def\H{\texttt{h}}
\def\E{\texttt{e}}

\newcounter{rown}

\renewcommand{\theequation}{\thesection.\arabic{equation}}

\begin{document}

\title{Basic quantizations of $D=4$ Euclidean, Lorentz, Kleinian and quaternionic $\mathfrak{o}^{\star}(4)$ symmetries
}
\author{A. Borowiec$^{1}$, J. Lukierski$^{1}$ and V.N. Tolstoy$^{2}$ \\
%EndAName
\\
$^{1}$Institute for Theoretical Physics, \\
University of Wroc{\l }aw, pl. Maxa Borna 9, \\
50--205 Wroc{\l }aw, Poland\\
\\
$^{2}$Lomonosov Moscow State University,\\
Skobeltsyn Institute of Nuclear Physics, \\
Moscow 119991, Russian Federation}

%\date{31.08-2017}%\date{}
\maketitle

\begin{abstract}
We construct firstly the  complete list of five quantum deformations of $D=4$ complex homogeneous orthogonal Lie algebra $\mathfrak{o}(4;\mathbb{C})\cong \mathfrak{o}(3;\mathbb{C})\oplus \mathfrak{o}(3;\mathbb{C})$, describing quantum rotational symmetry of  four-dimensional complex space-time, in particular we provide the corresponding universal quantum $R$-matrices. Further applying  four possible reality conditions we obtain all sixteen Hopf-algebraic quantum deformations for the real forms of $\mathfrak{o}(4;\mathbb{C})$: Euclidean $\mathfrak{o}(4)$, Lorentz $\mathfrak{o}(3,1)$, Kleinian $\mathfrak{o}(2,2)$ and quaternionic $\mathfrak{o}^{\star}(4)$. For $\mathfrak{o}(3,1)$ we only recall  well-known results obtained previously by the authors, but for other real Lie algebras (Euclidean, Kleinian, quaternionic) as well as for the complex Lie algebra $\mathfrak{o}(4;\mathbb{C})$ we present new results.

\end{abstract}

\tableofcontents

\setcounter{equation}{0}
\section{Introduction}
In recent years due to the efforts to construct models of quantum gravity characterized by non-commutative spacetime structures at Planckian distances \cite{Maj88}--\cite{Gar95}, the ways in which one can deform the algebras of spacetime coordinates and space-time symmetries became important. In noncommutative description of spacetime the numerical coordinates are replaced by noncommutative algebra, which is consistent with new type of uncertainty relations between the pairs of   operator-valued coordinates \cite{DFR95} called further DFR uncertainty  relation  relation. % \cite{Gar95}. 
This extension of noncommutativity into the spacetime sector describes
the limitations on spcetime localization measurements if the quantum gravitational background is present. It appears that during such procedure the high density of energy added by measurement leads to the creation of mini black holes, and one can show that below Planck distance $\lambda_P=10^{-33}m$ operationally the classical spacetime is not longer applicable. The quantum spacetime is effectively atomized, with lattice structure, and following the derivation in QM of Heisenberg algebra from Heisenberg uncertainty relations, one can deduce  from DFR uncertainty relation the noncommutativity of quantum spacetime.
 
Such noncommutative and/or discrete nature of quantum spacetime follows as well from loop quantum gravity (LQG) approach 
\cite{AshLew,Diana,Thi}, where the discretization is dynamical\footnote{We stress that in LQG spacetime lattice has  a dynamical origin, in particular it is not a way to regularize neither the QG action nor the QG functional integral in order to perform effectively the numerical calculations.},
based on the existence in LQG framework of minimal lengths, minimal area surfaces or minimal volume quanta. %\cite{Diana,Thi}.
In particular recently by applying LQG techniques to the quantum deformation of $D=3$ gravity with positive cosmological constant $\Lambda$ it has been shown \cite{JKG} that one gets the quantum symmetry of $U_q(\mathfrak{o}(4))$, where $\ln q\sim\Lambda$, in analogy with earlier results of \cite{LuRuNoTo} and \cite{Smolin} for Lorentz signature.

In this paper we shall consider the noncommutative structures as linked with the quantum groups, which are described as non-cocommutative  Hopf algebras \cite{Dr2,EtKazh,Wor87,ChPr94,Maj95,Klimyk,EtSch02}.
The deformed spacetime algebra is described as the irreducible representation (noncommutative Hopf algebra module) of quantum rotations algebra, with semidirect (smash) product  structure and build-in covariant action of quantum-deformed symmetry algebra on quantum noncommutative spacetime.
 
The aim of this paper is to provide the quantum Hopf algebras and universal $R$-matrices which are obtained by quantization of classical $r-$matrices classified recently in \cite{BLTnov15,LT17,BLT17}. In this paper we shall describe
the Hopf-algebraic deformations of any real four-dimensional rotational algebra given by the real form of $\mathfrak{o}(4;\mathbb{C})$. %\cite{BarutRacz77}.
Fortunately, all classical $r-$matrices presented in \cite{BLTnov15,BLT17} can be quantized by providing explicit formulae for coproducts,  antipodes and universal $R$-matrices. \footnote{For classification purposes we listed in  \cite{BLTnov15,LT17,BLT17} only the antisymmetric $r$-matrices. In the case of standard (or Drinfeld-Jimbo \cite{Dr2}) r-matrices one quantizes their symmetric Belavin-Drinfeld form \cite{BD82,BD84}, which satisfies CYBE and describes the leading order in the expansion of quantum $R$-matrix satisfying quantum Yang-Baxter equation \cite{Dr2,ChPr94,Maj95} For general formulae describing universal $R$-matrices see e.g. \cite{KhTo1}.}
We see therefore that the present paper provides the completion of the research program which we started in ref. \cite{BLTnov15,LT17,BLT17}.

The plan of our paper is the following:

In Sect.2 we shall present some generalities on quantization of 'infinitesimal' versions of quantum
deformations described by complex and real classical $r$-matrices, which provides the triangular and
nontriangular cases and further we present reality conditions for the universal $R$-matrices.
Further, in Sect. 3, we shall illustrate the quantization of   classical $r$-matrices by the explicit presentation, for $\mathfrak{sl}(2;\mathbb{C})$ case, of Jordanian and standard deformations.
%which we illustrate by considering real forms of two Drinfed-Jimbo [] and Jordanian [] deformation of $\mathfrak{sl}(2;\mathbb{C})$.

In Sect.~4 we shall recall $D=4$ complex Lie algebra $\mathfrak{o}(4;\mathbb{C})$ and its all four real forms. In particular, besides three real forms $\mathfrak{o}(4-k,k)$ ($k=0,1,2$) differing by the choice of signature, it should be added fourth quaternionic real form $\mathfrak{o}(2;\mathbb{H})\equiv\mathfrak{o}(2,1)\oplus \mathfrak{o}(3)\equiv\mathfrak{o}^*(4)$. We shall work mostly with the generators of four-dimensional complex rotations in Cartan-Weyl bases.

In Sect.~5 we quantize the full list of five classical $r$-matrices from \cite{BLTnov15,BLT17}, i.e. provide the complete list of all Hopf-algebraic deformations of $U_q(\mathfrak{o}(4;\mathbb{C}))$: two of them triangular, and remaining  three quasitriangular.\footnote{Only five $\mathfrak{o}(4;\mathbb{C})$ $r$-matrices are independent modulo $\mathfrak{o}(4;\mathbb{C})$ automorphism (see \cite{BLT17}).}
%can be obtained by considering the outer  exchanging the factors in the direct sum $\mathfrak{o}(3;\mathbb{C})\oplus \mathfrak{o}(3;\mathbb{C})$. }
In order to present the results in detail we shall calculate the coproducts, antipods and universal quantum $R$-matrices.
Further we specify all real forms of the quantum deformations of $\mathfrak{o}(4;\mathbb{C})$ described by $\star$-Hopf algebras. Following the standard  recipe (see e.g. \cite{KhTo1,LNR91,Wor91}) we assume that the $\star$-operation, defining  respective real form, acts on tensor product (coproduct) in unflipped way $(a\otimes b)^\star= a^\star\otimes b^\star$.
%The Lorentz $\mathfrak{o}(3,1)$ and Kleinian $\mathfrak{o}(2,2)$ cases will be considered rather briefly: 
We add that all quantum deformations of Lorentz algebra were already obtained earlier by the present authors \cite{BLT06,BLT08} and five out of eight  Kleinian $D=4$ real deformations can be obtained  from the complex $\mathfrak{o}(4;\mathbb{C})$ deformations listed in Sect.4 
simply by replacing the complex $\mathfrak{sl}(2;\mathbb{C})$ generators  by the real ones describing $\mathfrak{sl}(2;\mathbb{R})$ algebra. The Hopf-algebraic deformation of Euclidean $\mathfrak{o}(4)$ algebra some $\mathfrak{o}(2,2)$ deformations and quaternionic $\mathfrak{o}(2;\mathbb{H})$ case are the most important  because the obtained results are new. 

In Sect. 6 we shall present a brief outlook;  the paper contains also two appendices.
% in particular the perspective of extending the results to inhomogeneous and supersymmetric generators of $\mathfrak{o}(4;\mathbb{C})$ Lie algebras (see also \cite{BLMT12}--\cite{BLT15}). Finally we comment on the extensions of our considerations to $\mathfrak{o}(n;\mathbb{C})$, where $n=5,6.$

The quantum deformations of four-dimensional rotational symmetries presented in this paper can be applied at least in the following contexts:

 i) The deformed $D=4$ rotation groups with the signature $(+,+,-,-)$ (Kleinian case) describe  the deformed $D=3$ AdS symmetry and for Lorentz signature $(+,-,-,-)$ the $D=3$ dS quantum symmetries. If we introduce (A)dS radius $\Lambda$ and the re-scaling of three rotations $M_{1k}\rightarrow \tilde M_{1k}=\Lambda P_k$ ($P_k$ describes curved (A)dS momenta, $k=1,2,3$), by suitable quantum Wigner-In\"{o}n\"{u} contraction \cite{CeGiSoTa,LuRuNoTo} one can get various $\kappa$-deformed $D=3$ Poincar\'{e} algebras.

ii) The knowledge of classical $r-$matrices permits to introduce explicitly the action of deformed (super)string models, described by so-called YB (Yang-Baxter) sigma models \cite{Klim02}--\cite{BKLSY15}. The quantum deformations presented in this paper can be applied to the description of the YB deformation of principal $\mathfrak{o}(4-k,k)$  $\sigma-$models ($k=0,1,2$) as well as to the coset sigma models with noncommutative target space, described by the deformed cosets $\frac{\mathfrak{o}(4-k,k)}{\mathfrak{o}(3-k,k)}$. It appears that such deformed  $\mathfrak{o}(4-k,k)$ group or their
%$\mathfrak{o}(4-k,k)$ 
coset manifolds can appear as parts of internal symmetry target spaces obtained by the reduction to $D=4$ of deformed $D=10$ Green--Schwarz superstrings.

iii) The classical $r$-matrices and their quantizations provide a powerful algebraic tool in description of integrable models and provide effective methods for studying their multihamiltonian systems \cite{VDr,MBl,StS}. In particular, the methods of noncommutative geometry permits to consider as well the Hamiltonian theories over the noncommutative rings \cite{Dorfman,Fokas} and their integrability conditions.

iv) Eight quantum deformations of $\mathfrak{o}(2,2)$ presented in the paper provide the set of finite $D=2$ quantum conformal algebras, with six generators, which in general case cannot be factorized into a sum of "`left"' and "`right"' ($X_\pm=X_1\pm X_0$) $D=1$ quantum deformed conformal algebras. It is interesting to study  which $\mathfrak{o}(2,2)$ deformations  presented in the paper can be consistently extended to infinite-dimensional quantum groups, describing new classes of deformed $D=2$ infinite-dimensional conformal Virasoro algebras.

v) For various real forms of quantum-deformed $\mathfrak{o}(4;\mathbb{C})$ groups one can obtain corresponding four-dimensional spacetime with different signatures (see e.g. \cite{Rita17}). With all Hopf-algebraic deformations of 
$\mathfrak{o}(4;\mathbb{C})$ which will be presented in this paper one can obtain the complete list of quantum spacetimes with signatures $(4,0)$, $(3,1)$ and $(2,2)$.

Further remarks related with the applications of  quantum deformations considered in this paper we shall present also in Sect. 6.

\section{Quantizations of complex and real Lie algebras: general remarks}

\subsection{From classical $r$-matrices to quantum universal $R$-matrices}

It is  known that formulated by  Drinfeld \cite{Dr2} the quantization problem   
%(V. G. Drinfeld [in Quantum groups (Leningrad, 1990)) 
of Lie bialgebras has been answered  by Etingof and Kazhdan \cite{EtKazh}: to each Lie bialgebra one can associate a quantized enveloping algebra supplemented with Hopf algebra structure.  Unfortunately, their proof is not constructive and the methods of explicit quantizations are known  only in specific situations, as e.g. Drinfeld-Jimbo quantization of semi-simple Lie algebras and twist quantization in the triangular case (when twist tensor can be constructed explicitly). We shall show however that the known quantization techniques are sufficient  for finding all explicit non-isomorphic quantizations of the enveloping algebra $\mathfrak{o}(4;\mathbb{C})$ and their real form.

Principal tool for the classification of quantum deformations is provided by the classical $r$-matrices \cite{BD82,BD84,Sem-T-Sh83}  which determine  coboundary Lie bialgebra \footnote{With the cobracket given by the commutator $\delta_r(x)=[x\otimes 1+1\otimes x,r]$, see e.g. \cite{ChPr94,Maj95,EtSch02}.}
structures. Quantization procedure od bialgebras leads to   the construction of quantum-deformed associative and coassociative Hopf-algebras \cite{Dr1} %quantum enveloping algebra 
and determine the corresponding universal (quantum) $R-$matrices \cite{ChPr94,Maj95,KhTo1}.  

For semi-simple Lie algebras, due to the classical Whithead lemma, all bialgebras are coboundary.  In such a case there is one-to-one correspondence between the Lie bialgebra structure and the corresponding  classical $r$-matrix given as the skewsymmetric  element $r\in\mathfrak{g}\wedge\mathfrak{g}$ satisfying the classical 
(homogenous or inhomogenous) YB  equation:
\begin{eqnarray}\label{crm1}
[[r,\,r]]\!\!&=\!\!& t\,\Omega , \quad t\,\in\mathbb{C}~.
\end{eqnarray}
with $[[\cdot,\cdot]]$ denoting  Schouten bracket 
\begin{equation}\label{i4}
    [[r,r]]\equiv [r_{12},r_{13}+r_{23}]+[r_{13},r_{23}].
\end{equation}
where %$[[r,r]]$ denotes Schouten brackets  and
$r_{12}=r^{(1)}\otimes r^{(2)}\otimes 1\in \mathfrak{g}\otimes\mathfrak{g}\otimes\mathfrak{g}$ etc.
For skew-symmetric 2-tensor monomials $x\wedge y=x\otimes y-y\otimes x$ and $u\wedge v$ ($x,y,u,v\in\mathfrak{g}$) the explicit  formula for Schouten brackets reads\footnote{For general elements  $r_{1}, r_2\in \mathfrak{g}\wedge\mathfrak{g}$ one can extend (\ref{schouten}) by  bilinearity.}
\begin{eqnarray}\label{schouten}
\begin{array}{rcl}
[[x\wedge y,\,u\wedge v]]\!\!&:=\!\!&x\wedge\bigl([y,u]\wedge{v}+u\wedge[y,v]\bigr)
\\[4pt]
&&-y\wedge\bigl([x,u]\wedge{v}+u\wedge[x,v]\bigr)
\\[5pt]
\!\!&\phantom{:}=\!\!&[[u\wedge v,\,x\wedge y]]
\end{array}
\end{eqnarray}
where the three-form $\Omega$ is the $\mathfrak{g}$-invariant element in $\mathfrak{g}\wedge\mathfrak{g}\wedge\mathfrak{g}$, i.e.
\begin{equation}\label{i8a}
 ad_x\Omega\equiv[x\otimes 1\otimes 1+1\otimes x\otimes 1+1\otimes 1\otimes x, \Omega]=0 ,\quad x\in\mathfrak{g}.
\end{equation}
The complex Lie bialgebra is described by a pair $\mathsf{g}\equiv(\mathfrak{g}, r)$ consisting of complex Lie algebra $\mathfrak{g}$
and skew-symmetric classical $r$-matrix $r$ satisfying the equation (\ref{crm1}).
One can distinguish two cases (cf. e.g. \cite{ChPr94,Maj95,EtSch02}):

A) If in (\ref{crm1}) $t= 0$, one gets the so-called triangular or non-standard case with vanishing Schouten brackets describing homogenous classical Yang-Baxter equation (denoted as CYBE). In such a case the $r$-matrix can be rescaled arbitrarily without changing the corresponding bialgebra structure.
The  triangularity is preserved by  Lie algebra homomorphisms and can be reduced to non-degenerate case on 
the Borel subalgebra.

B) If $t\neq 0$, eq. (\ref{crm1}) describes so-called non-triangular (quasitriangular)  classical $r$-matrix, satisfying inhomogenous  or modified classical Yang-Baxter equation (mCYBE). In such case  one can introduce $ r_{BD}\in \mathfrak{g}\otimes \mathfrak{g}$  called Belavin-Drinfeld form of the $r$-matrix satisfying CYBE, such that $r= r_{BD}- r_{BD}^\tau$ ($(x\otimes y)^\tau=y\otimes x$ is the flip operation) and the symmetric element $r_{BD}^s\equiv r_{BD} +r_{BD}^\tau$ which is ad-invariant.
\footnote{The symmetric part $r_s$ is $\mathfrak{g}$-invariant and in the case of semi-simple algebra is related to the so-called split Casimir (non-degenerate Cartan-Killing form).}
In general, the initial skew-symmetric $r$-matrix is not scale invariant nor preserved by a Lie algebra homomorphisms.
It is remarkable that Belavin-Drinfeld $r$-matrices for simple Lie algebras has been fully classified by means of so-called Belavin-Drinfel triples in \cite{BD84}.

Quantization of (complex) Lie bialgebra leads to quantum groups in Drinfeld sense \cite{Dr2} with the Hopf algebra structure 
supplementing the complex deformed universal enveloping algebra $U(\mathfrak{g})$ (in general one needs its topological $\xi$-adic extension $U_\xi(\mathfrak{g}) \equiv U(\mathfrak{g})[[\xi]]$ formulated also for multiparameter deformation, i.e. $\xi\rightarrow (\xi_1,\ldots,\xi_k)$, see e.g. \cite{Dr2,ChPr94,Maj95,Klimyk}).
According to the cases A), B) indicated above, there are two ways of introducing  quantum-enveloping algebra  what will be described shortly below. Before we would like to focus  our attention on the quantum universal $R$-matrix as an important  byproduct of the quantization procedure.  
\footnote{The importance of quantum $R$-matrices follows from their applications as solutions
of qYBE in various branches of theoretical physics
e.g. conformal field theory, statistical mechanical models, and in mathematics, e.g. for description of link invariants.}

The universal $R$-matrix is  an invertible element of $U_\xi(\mathfrak{g})\otimes U_\xi(\mathfrak{g})$
which provides the flip $\tau$ of the noncocommutative coproduct
$\tau:\Delta_\xi\rightarrow \Delta^\tau_\xi$ 
%($\Delta^{}_\xi=\Delta^{(1)}_\xi\otimes\Delta^{(2)}_\xi; \Delta^{\tau}_\xi\equiv\tau\circ\Delta^{}_\xi=\Delta^{(2)}_\xi\otimes\Delta^{(1)}_\xi$)
given by the following similarity transformation
\begin{equation}\label{i2}
    \Delta^\tau_\xi(\cdot)= R \,\Delta_\xi(\cdot) \,R^{-1}
\end{equation}
The universal $R-$matrix describes quantum group (see \cite{Dr1}) if it satisfies quasitriangularity conditions
\begin{equation}\label{i3a}
    (\Delta^{}_\xi\otimes id)R =R_{12} R_{23}  ,\qquad
    (id\otimes\Delta^{}_\xi)R =R_{13} R_{12} 
\end{equation}
where $R =R^{(1)} \otimes R^{(2)} $ and $R_{12} =R^{(1)} \otimes R^{(2)} \otimes 1$, etc..
The  properties (\ref{i2})--(\ref{i3a})
%determine so-called quasi-triangular structure
%on the underlying Hopf algebra $(U_\xi(\mathfrak{g}), \Delta_\xi, S_\xi,\epsilon)$ which
imply in $U_\xi(\mathfrak{g})\otimes U_\xi(\mathfrak{g})\otimes U_\xi(\mathfrak{g})$ quantum Yang-Baxter equation (qYBE) in the form 
 \begin{equation}\label{i3b}
     R_{12} R_{13} R_{23} =R_{23}  R_{13} R_{12} 
\end{equation}
 as well as the following normalization conditions %(see e.g. \cite{Klimyk} p.245)
\begin{equation}\label{i3c}
(\epsilon\otimes id)R =(id \otimes\epsilon)R =1,%\quad (S_\xi \otimes id)R(\xi)=R^{-1}(\xi), \quad
%(S_\xi \otimes id)R^{-1}(\xi)=R(\xi), \quad
\end{equation}
%$(S_\xi \otimes S_\xi)R(\xi)=R(\xi)$,
where $\epsilon$  denotes a counit. 

In fact, the same properties (\ref{i2})--(\ref{i3c}) are satisfied by another 
universal $R$-matrix, which is $(R^\tau)^{-1}$. Therefore one can distinguish two case:

i) the element $Q_R=R R^\tau=1$ is trivial

ii) $Q_R\neq 1$ is  non-trivial\\
\noindent It turns out that the first case corresponds to the triangular or twist quantization case while the second characterizes the non-triangular case. In order to describe their difference let us expand the $R$-matrix (\ref{i2}) in the powers of the deformation parameter $\xi$, entering linearly in the definition of classical $r$-matrix \footnote{The parameter $\xi$ in $o(\xi^2)$ should be replaced by $(\xi_1,\ldots,\xi_k)$ in the case of multiparameter deformation, i.e. for multiparameter classical $r$-matrix which is
linear in $\xi_i\ (i=1,\dots,k)$ the expansion (\ref{i1}) is up to any quadratic term in $\xi_i$.}
\begin{equation}\label{i1}
    R(\xi)= 1\otimes 1 + \tilde r + O(\xi^2)
\end{equation}
From (\ref{i1}) and (\ref{i3b}) it follows that the element $\tilde r\in \mathfrak{g}\otimes\mathfrak{g}$  satisfies  classical Yang--Baxter equation (CYBE) \cite{ChPr94,Maj95}. In triangular case one has $\tilde r+\tilde r^\tau=0$, i.e. $\tilde r$ is skew-symmetric and can be identified with the classical $r$-matrix satisfying (\ref{crm1})  with $t=0$. In the second (non-triangular, see ii)) case, $\tilde r$ is not skew-symmetric, satisfies CYBE and takes the  Belavin-Drinfeld form of $r$-matrix, i.e. $\tilde r=r_{BD}$.

The classical $r$-matrices describe the infinitesimal version of quantum deformed Lie-algebraic symmetries; the quantum deformation parameterized by an arbitrary (formal) deformation parameter $\xi$ determines Hopf-algebraic quantization and universal $R-$matrix . 
%We see that either the classical $r$-matrix $r=r^{(1)}\wedge r^{(2)}\in\mathfrak{g}\wedge\mathfrak{g}$ or its Belavin-Drinfeld form $r_{BD}=r_{BD}^{(1)}\otimes r_{BD}^{(2)}\in\mathfrak{g}\otimes\mathfrak{g}$ does appear as the first order term in the power series expansion (\ref{i1}).% in the deformation parameter $\xi$.
 
In general case it is not known  how to obtain the universal $R$-matrix from the solutions of (\ref{i4}); however for canonical Belavin--Drinfeld nontriangular $r-$matrices  \cite{BD82} the explicit formula for universal $R-$matrices is well-known (see e.g. \cite{KhTo1}). It is worth noticing that in contrast to the triangular case, the non-triangular one provides two different quantum $R$-matrices: $R(\xi)$ and $R^\tau(\xi)^{-1}$. The element $Q(\xi)\equiv R^\tau(\xi)R(\xi)=1+(r+r^\tau) + O(\xi^2)$ is called a quantum Killing form since its first order term, if not degenerate, defines a classical Cartan-Killing form on $\mathfrak{g}$.
We would like to add that the skew-symmetric classical $r$-matrices are sufficient for classification \footnote{They are more tractable, since $dim\,(\mathfrak{g}\wedge\mathfrak{g})<dim\,(\mathfrak{g}\otimes\mathfrak{g})$.} as well as for the description of correspondence  with classical Lie-Poisson groups. 
%In the present paper non-skewsymmetric $r$-matrices will be helpful in order to distinguish between two different quantum reality conditions inspired by universal $R$-matrix properties.

\subsection{Reality conditions providing the quantizations of real bialgebras} 
  
We remind that a real Lie algebra structure $(\mathfrak{g}, \divideontimes)$ can be introduced by adding  an antilinear involutive (Lie algebra) anti-automorphism $\divideontimes : \mathfrak{g}\rightarrow\mathfrak{g}$ ($\divideontimes$-operation, conjugation) acting on the complex Lie algebra $\mathfrak{g}$. It relies on finding Lie algebra basis with real structure constants for which $\divideontimes$-operation is anti-Hermitian (i.e. $x^\divideontimes=-x$)
\footnote{In a case of Hilbert space realization this condition leads to operators with imaginary spectrum. For this reason some authors do prefer instead Hermitian generators and imaginary structure constants as representing real Lie algebras.}. Subsequently, the real coboundary Lie bialgebra can be considered as a triple $(\mathsf{g},\divideontimes)\equiv(\mathfrak{g}, \divideontimes, r)$, where the skew-symmetric element $r$ is assumed to be anti-Hermitian, i.e. 
%\footnote{Such conditions have to be imposed on the $r$-matrix together with its (formal or numerical) parameters.}
\begin{equation}\label{rc_1}
	r^{\divideontimes\otimes\divideontimes}=-r =r^\tau.
\end{equation}
Such conditions lead to the suitable reality conditions for
%have to be imposed on the $r$-matrix together with its (formal or numerical) 
parameters, what is particularly important in the inhomogenous case ($t\neq 0$), due to the lack of scale invariance.

The $\divideontimes$-operation extends, by  the  property $(ab)^\divideontimes=b^\divideontimes a^\divideontimes$ (i.e. as an antilinear antiautomorphism),  to the enveloping algebra $U_{}(\mathfrak{g})$, as well as to quantized enveloping algebra, making both of them associative $\divideontimes$-algebras.
The real Hopf-algebraic structure represented on quantized enveloping algebra $U_{q_{}}(\mathfrak{g})$ by $\divideontimes$-involution is  defined by the following %compatibility 
conditions for  coproducts and antipodes (see also \cite{LNR91,Wor91})
\begin{equation}\label{d-1}
\Delta_{q_{}}(a^\divideontimes)\;=\;(\Delta_{q_{}}(a))^\divideontimes,\quad\;\; S_{q_{}}((S_{q_{}}
(a^\divideontimes))^{\divideontimes})\;=\;a,\quad\;\epsilon(a^\divideontimes)=\epsilon(a)^*\quad(\forall a\in U_{q_{}}(\mathfrak{g})~.
\end{equation}
where the  $\divideontimes$-involution on the tensor product (\ref{d-1}) acts as follows
\footnote{In other words a real Hopf algebra is identified with a $\star$-Hopf algebra which is a complex Hopf algebra equipped with an additional star operation making the algebraic sector into $\star$-algebra and coalgebraic sector satisfying  (\ref{d-1}).}
\begin{equation}\label{d-1b}
(a\otimes b)^\divideontimes = a^\divideontimes\otimes b^\divideontimes%%\qquad(\divideontimes-{\rm direct})~.
\end{equation}
%Following the standard approach (see e.g. \cite{Maj95}) 
%It is possible  to  
One can get (\ref{i2}) and (\ref{d-1}) compatible and consistently  defined
quasitriangular  $\divideontimes$-Hopf algebras by imposing   two distinct reality constraints on the universal $R$-matrix (see e.g. \cite{Maj95}):

a) $R^{\divideontimes\otimes\divideontimes}=R^\tau$ ($R$ is called real);

b) $R^{\divideontimes\otimes\divideontimes}=R^{-1}$ and the corresponding quantum $R$-matrix is $\divideontimes$-unitary ($R$ is called antireal).

Particularly, in the triangular case, %described by twist quantization (see below), 
due to the identity $R^\tau=R^{-1}$, the conditions a) and b) 
are the same. In non-triangular case  ($R^\tau\neq R^{-1}$), the second universal R-matrix $(R^\tau)^{-1}$ satisfies  the same reality constraints 
\footnote{It should be observed that the presence of these two universal $R$-matrices may help to obtain finite contraction limits (see e.g. \cite{Beisert17})}.

It should be noted that for any element $\tilde r\in\mathfrak{g}\otimes\mathfrak{g}$ satisfying CYBE, $\tilde r^{\divideontimes\otimes\divideontimes} $ %
satisfies again CYBE. Therefore, one can distinguish two cases:

i) the classical $r$-matrix $r=\tilde r$ corresponding to the universal $R$-matrix (cf. \ref{i1}) is skew-symmetric; then $r$ should be anti-Hermitian and satisfy the relation  (\ref{rc_1}). 
%$$(\xi r)^{\divideontimes\otimes\divideontimes}\equiv \xi^*\,r^{\divideontimes\otimes\divideontimes} =\xi\,r^\tau
%\equiv-\xi\,r.$$ 
%%This corresponds to the triangular (non-standard) case with quantization procedure leading to the unitary twist deformation (\ref{i6}).
%Thus for $r$ Hermitian ($r=r^{\divideontimes\otimes\divideontimes}$), the parameter $\xi$ must be imaginary.

ii) if the element $\tilde r$ is not skew-symmetric this corresponds to the non-triangular case; then $\tilde r^{\divideontimes\otimes\divideontimes}=\tilde r^\tau$
\footnote{This condition can be rewritten as $\tilde r^{\tau(\divideontimes\otimes\divideontimes)}=\tilde  r$, where $(a\otimes b)^{\tau(\divideontimes\otimes\divideontimes)}=b^\divideontimes \otimes a^\divideontimes$ denotes so-called flipped conjugation (cf. \cite{LNR91,BLT08} and formula (\ref{d-1f}) below).}
 for $R$ real and $\tilde r^{\divideontimes\otimes\divideontimes}=-\tilde r$ for $R$ antireal. It is easy to check that in any case the skew-symmetric part of $\tilde r$ remains anti-Hermitian, i.e.
 $(\tilde r-\tilde r^\tau)^{\divideontimes\otimes\divideontimes}=-(\tilde r-\tilde r^\tau)$.   \medskip\\
 %In what follows we shall need the following obvious fact.\medskip\\
It is easy to show that twisting of real form of quasitriangular Hopf algebra by unitary twist leads again to real quasitriangular Hopf algebra. More precisely,  
%{\bf Lemma 1}:%\textit{
if $(\mathcal{H}, \Delta, S, \epsilon, R, \star)$ is a quasitriangular $\star$-Hopf algebra with $R$ being real (resp. antireal) universal $R$-matrix, then for any unitary, normalized 2-cocycle twist $F=(F^{-1})^{\star\otimes\star}\in \mathcal{H}\otimes\mathcal{H}$ the quantized algebra $(\mathcal{H}, \Delta_F, S_F, \epsilon, R_F, \star)$ is a quasitriangular $\star$-Hopf algebra such that $R_F=F^\tau R F^{-1}$ is real (resp. antireal).
This property will be used in the consideration of some cases of chain quantization of $\mathfrak{o}(4;\mathbb{C})$
(see Sect. 5).

\section{The basic $\mathfrak{sl}(2;\mathbb{C})$ example: complex and real Lie bialgebra and their quantizations}

\subsection{Complex and real bialgebras }

We recall that classical $r$-matrices, providing Lie bialgebra structure of a given Lie algebra as well as quantum deformations of the corresponding enveloping algebra, are classified up to the isomorphisms;
%(i.e. isomorphisms preserving the structure constants).
in particular for real Lie algebras one should use the isomorphisms  preserving reality condition. 
\footnote{From now on all Lie algebras and bialgebras are real if not indicated otherwise.}
Fixing the basis (structure constant) one deals with Lie algebra automorphisms.
For simple Lie algebras these are   (modulo discrete automorphisms) the internal automorphisms  generated by the adjoint actions of the Lie algebra upon itself.   The important example of such classification for  real forms of $\mathfrak{o}(4;\mathbb{C})$ has been investigated in  \cite{BLTnov15,LT17,BLT17}. 
%Thus complex coboundary Lie bialgebra can be considered as a pair $(\mathfrak{g}, r)$ consisting
%of complex Lie algebra and a solution of CYBE/mCYBE. In such approach real Lie bialgebra is a triple $(\mathfrak{g}, \divideontimes, r)$ consisting of complex Lie algebra, $\divideontimes$-involution (real form) and a Hermitian/anti-Hermitian solution of CYBE.\medskip\\

It is well known that for the complex Lie algebra $\mathfrak{sl}(2;\mathbb{C})\cong\mathfrak{o}(3;\mathbb{C})$
there exists up to $\mathfrak{sl}(2,\mathbb{C})$ automorphisms two solutions of mCYBE
\footnote{There are only two orbit types under the action of $\mathfrak{o}(3;\mathbb{C})$ in $\mathbb{C}^3$: null and non-null, see Appendix A.}, namely Jordanian $r_{J}^{}$ (triangular, called also non-standard) and the standard one 
$r_{st}^{}$ (non-triangular):
\begin{eqnarray}\label{rm1a}
r_{J}^{}\!\!&=\!\!&\xi\,E_{+}\wedge H,\quad[[r_{J}^{},r_{J}^{}]]\;=\;0,
\\[3pt]\label{rm2}
r_{st}^{}(\gamma)\!\!&=\!\!&\gamma E_{+}^{}\wedge E_{-}^{},\quad[[r_{st}^{},r_{st}^{}]]\;=\;{\gamma}^{2}\Omega,
\end{eqnarray}
where we use the Cartan--Weyl (CW) basis 
\begin{equation}\label{rm3}
    [H,E_{\pm}]\,=\,\pm E_{\pm},\qquad [E_{+},E_{-}]\,=\,2H .
\end{equation}
%Two comments are now in order. 
In (\ref{rm1a}) the parmeter $\xi$ can be replaced by $\xi=1$ due to the scale invariance of CYBE.
In the standard case (\ref{rm2}) the non skew-symmetric counterpart of $r_{st}$ satisfies CYBE if it takes 
the Belavin-Drinfeld form
\begin{eqnarray}\label{rm5}
r_{BD}^{}(\gamma)\!\!&=\!\!&\gamma\bigl(E_{+}\otimes E_{-}+H\otimes H\bigr)
%+\bar\gamma\bigl(\bar E_{+}\otimes \bar E_{-}+\bar H\otimes \bar H\bigr)
\end{eqnarray}
Its symmetric part, described by $\mathfrak{sl}(2;\mathbb{C})$ bilinear split Casimir
$$E_{+}\otimes E_{-}+ E_-\otimes E_+ +2H\otimes H$$ is an invariant element in $\mathfrak{sl}(2,\mathbb{C})\otimes \mathfrak{sl}(2,\mathbb{C})$ and determines the Cartan-Killing form.

We recall that the (complex) simple Lie algebra $\mathfrak{o}(3; \mathbb{C})\cong\mathfrak{sl}(2; \mathbb{C})$ has up to $\mathfrak{sl}(2; \mathbb{C})$-isomorphisms two real forms: compact  $\mathfrak{o}(3)\cong\mathfrak{su}(2)$ 
%\footnote{From now on all Lie algebras and bialgebras are real if not indicated otherwise.} 
and noncompact $\mathfrak{o}(2,1)\cong\mathfrak{su}(1,1)\cong  \mathfrak{sl}(2)$. 
%The above notations are related with different (faithful) matrix realizations satisfying corresponding reality conditions. 
It is known, see e.g. \cite{LT17}, that with these  two real forms there are linked four real Lie bialgebras, one compact and three noncompact ones, which can be expressed in
$\mathfrak{su}(1,1)\cong \mathfrak{sl}(2)$ or $\mathfrak{o}(2,1)$ bases  (see also Appendix A).
 
% as the triples indicated above:
The unique compact real bialgebra one can write in $\mathfrak{su}(2)$ basis (cf. \cite{Maj95,Klimyk})
\footnote{We are working in the Cartan Weyl basis and different reality conditions, cf. \cite{LT17}.},
\footnote{Further we shall use specific notation  in order to distinguish
between real Lie algebras and bialgebras, e.g. $\mathsf{sl}_\gamma(2)$ denotes
the triple $(\mathfrak{sl}(2;\mathbb{C}),\#, r_{st}(\gamma))$} satisfying reality conditions ($\divideontimes=\dag$)
\begin{eqnarray}\label{rm6-1} H^{\dagger}\!\!&=\!\!&H,\qquad E_{\pm}^{\dagger}\;=\;E_{\mp},\quad r_{st}(\gamma)\, ,\,\gamma\in \mathbb{R} \quad \;\;(\mathsf{su}_\gamma(2)\;\; {\rm{standard\, bialgebra}})
\end{eqnarray}
From three noncompact inequivalent real bialgebras we choose to write one in $\mathfrak{su}(1,1)$ basis ($\divideontimes=\#$)
\begin{eqnarray}\label{rm6-2}
H^{\#}\!\!&=\!\!&H,\quad\;E_{\pm}^{\#}\;=\;-E_{\mp},\quad\;  r_{st}(\gamma)\, ,\,\gamma\in \mathbb{R}
\quad\;\;(\mathsf{su}_\gamma(1,1)
\;\; {\rm{standard\, bialgebra}})
\end{eqnarray}
and remaining two in $\mathfrak{sl}(2)$ basis ($\divideontimes=\star$)
\begin{eqnarray} \label{rm6-3} H^{\star}\!\!&=\!\!&-H,\quad\;E_{\pm}^{\star}\;=\;-E_{\pm},\quad\;\; r_{st}(\gamma)\, ,\,\gamma\in \imath\mathbb{R} \quad\;\;(\mathsf{sl}_\gamma(2)\;\; {\rm{standard\, bialgebra}})
\end{eqnarray}
\begin{eqnarray}\label{rm6-4}
 H^{\star}\!\!&=\!\!&-H,\quad\;E_{\pm}^{\star}\;=\;-E_{\pm},\quad\;\; r_{J}\quad %\Leftrightarrow\gamma\in \imath\mathbb{R}
 \qquad\quad (\mathsf{sl}_J(2)\;\qquad {\rm{nonstandard\,\, bialgebra}})
 \end{eqnarray}
First three $r$-matrices (\ref{rm6-1})-(\ref{rm6-3}) are standard (non-triangular) while the last  (\ref{rm6-4}) is  
Jordanian (triangular) without multiplicative parameter because it has been rescaled to $1$ by suitable $\mathfrak{sl}(2)$-automorphism. The first and second bialgebra depends on real parameter, and the third one is multiplied by purely imaginary parameter (it is antireal). We stress that however $\mathsf{su}(1,1)$ and  $\mathsf{sl}(2)$ are isomorphic with  real Lie algebra $\mathfrak{o}(2,1)$, they are not isomorphic as  real Lie  bialgebras (cf. \cite{LT17}).

In formulae (\ref{rm6-1})-(\ref{rm6-3}) there are used for the complex CW basis (\ref{rm3}) three different  reality conditions defining $\mathfrak{su}(2)$, $\mathfrak{su}(1,1)$ and  $\mathfrak{sl}(2)$ real algebras. Because 
$\mathfrak{o}(2,1)\cong\mathfrak{su}(1,1)\cong  \mathfrak{sl}(2)$, the involutions $\#$ and  $\star$ (see (\ref{rm6-2})-(\ref{rm6-3})) can be identified and related with the reality condition defining $\mathsf{o}(2,1)$ as real form of $\mathsf{o}(3;\mathbb{C})$. Indeed, the $\mathfrak{su}(1,1)$ real basis $(H', E_\pm')$ and $\mathfrak{sl}(2)$ real bases $(H, E_\pm)$ can be related by the following 
linear complex $\mathsf{sl}(2;\mathbb{C})\cong\mathfrak{o}(3; \mathbb{C})$ automorphism
\begin{eqnarray}\label{jl1}
\begin{array}{rcl}
H' = \displaystyle-\frac{\imath}{2}\big(E_{+}-E_{-}\big) ,\quad
%\\[7pt]
E_{\pm}'= \displaystyle\mp\imath H+\frac{1}{2}\big(E_{+}+E_{-}\big).
\end{array}
\end{eqnarray}
One can use the complex Cartesian basis $I_k\in \mathfrak{o}(3; \mathbb{C})$  ($k=1,2,3$)
%(being at the same time a basis for its real compact form $\mathfrak{o}(3)\cong\mathfrak{su}(2)$)
\begin{eqnarray}\label{jl2}
\begin{array}{rcl}
[I_{i},\,I_{j}]\!\!&=\!\!&\varepsilon_{ijk}I_{k}
\end{array}%
\end{eqnarray}
which is antireal for the real compact form $\mathfrak{o}(3)\cong\mathfrak{su}(2)$
 \begin{eqnarray}\label{jl3b}
I_{i}^{\dag}\!\!&=\!\!&-\,I_{i}^{}\quad(i=1,2,3)\quad{\rm{for}}\;\;\mathfrak{o}(3).
\end{eqnarray}
%Further, the relation between $\mathfrak{su}(1,1)$ and $\mathfrak{sl}(2)$ bases becomes more transparent. 
For both cases
$\mathfrak{su}(1,1)$ and $\mathfrak{sl}(2)$ the reality condition  in Cartesian basis takes the same 
$\mathfrak{o}(2,1)$ form
 \begin{eqnarray}\label{jl3}
I_{i}^{\star}\!\!&=\!\!&(-1)^{i-1}I_{i}^{}\quad(i=1,2,3)\quad{\rm{for}}\;\;\mathfrak{o}(2,1).
\end{eqnarray}
%(which differs from the compact $\mathfrak{su}(2)$ conjugation: $I_i^\ddagger=-I_i$).
%with $I_2$ as noncompact (anti-Hermitian) generator.
One can relate the $\mathfrak{su}(1,1)$ and $\mathfrak{sl}(2)$ bases  with the Cartesian $\mathsf{o}(2,1)$ generators satisfying the same reality  condition (\ref{jl3}) by the following formulae
\begin{eqnarray} 
&&\begin{array}{rcl}\label{jl4}
&&H'\,:=\,\imath I_{2},\qquad\qquad E_{\pm}'^{}\,:=\,\imath I_{1}\pm I_{3},
%\\[4pt]
%&&[H,E_{\pm}^{}]\,=\,\pm E_{\pm}^{},\quad[E_{+}^{},E_{-}^{}]\,=\,2H\end{array}
\qquad{\rm{for}}\;\;\mathfrak{su}(1,1),%
\\[4pt]
%&&\begin{array}{rcl}\label{pr8}
&&H\,:=\,\imath I_{3},\qquad\qquad E_{\pm}\,:=\,\imath I_{1}\mp I_{2}, 
%\\[4pt]
%&&[H',E_{\pm}']\,=\,\pm E_{\pm}',\quad[E_{+}',E_{-}']\,=\,2H'\end{array}
\qquad{\rm{for}}\;\;\mathfrak{sl}(2,\mathbb{R}).\end{array}%
\end{eqnarray}
Both CW bases $\{E_{\pm}',H'\}$ and $\{E_{\pm},H\}$ have %the same commutation relations but they have 
different reality properties which follow from the same reality condition (\ref{jl3})
\begin{eqnarray} 
\begin{array}{rcl}\label{jl5}
&&{H'}^{\star}\;=\;H',\qquad\;\; {E_{\pm}'}^{\star}=-E_{\mp}'^{}
%\end{array}
\qquad{\rm{for}}\;\;\mathfrak{su}(1,1),
\\[3pt]
%&&\begin{array}{rcccl}\label{pr10}
&&{H}^{\star}\;=\;-H,\qquad{E_{\pm}}^{\star}=-E_{\pm}
%\end{array}
\qquad{\rm{for}}\;\;\mathfrak{sl}(2;\mathbb{R}),\end{array}
\end{eqnarray}
%where the conjugation ($^{\dag}$) is the same as in (\ref{jl3})\footnote{
It should be noted that in the case of $\mathfrak{su}(1,1)$ the Cartan generator $H'$ is compact while for the case $\mathfrak{sl}(2)$ the generator $H$ is noncompact, what also explains the difference between $\mathfrak{su}(1,1)$ and $\mathfrak{sl}(2)$ CW basis. 
In this way the involutions (\ref{jl3}) and  (\ref{rm6-2}) -- (\ref{rm6-3}) are identified  (it can be checked that the relations (\ref{jl1}) and (\ref{jl4})  are consistent).
%and we repeat that the difference between $\mathfrak{su}(1,1)$ and $\mathfrak{sl}(2)$ CW bases lies in compact and noncompact assignments of the Cartan generator $H$. 
Concluding, it is 
sufficient for $\mathfrak{sl}(2; \mathbb{C})$ to introduce only two involutions: defining $\mathfrak{o}(3)\cong\mathfrak{su}(2)$
and $\mathfrak{o}(2,1)\cong\mathfrak{su}(1,1)\cong  \mathfrak{sl}(2)$.
In fact the formulae (\ref{jl4}) can be used for the introduction of Cartesian basis in all classical $\mathfrak{o}(2, 2)$ and $\mathfrak{o}^*(4)$
$r$-matrices containing the $\mathfrak{su}(1,1)$ and $\mathfrak{sl}(2)$ sectors.

 %$$\mathsf{su}_q(2),  \mathsf{sl}_q(2), \mathsf{su}_q(1,1), \mathsf{sl}_J(2),$$

In the next subsection we shall describe explicitly the quantization of the complex bialgebras (\ref{rm1a}) and (\ref{rm2}). In order to obtain the quantization of the bialgebras listed in (\ref{rm6-1}) - (\ref{rm6-4}) one should insert the generators $(H, E_\pm)$ satisfying the respective reality condition and impose the suitable restriction on the parameter $\gamma$. Standard deformation of simple Lie algebra is given by the explicit algorithm introduced firstly by Drinfeld and Jimbo.
Non-standard  quantum deformation of $\mathsf{g}\equiv(\mathfrak{g}, r)$, where $r$ a skew symmetric solution of CYBE, is obtained by employing the 2-cocycle  Drinfeld twist element $ F \in U(\mathfrak{g}) \otimes
U(\mathfrak{g})$ which remains unchanged the algebra and modifies the coproduct $\Delta$ and antipode $S$ as follows (see e.g. \cite{ChPr94,Maj95}):
\begin{equation}\label{blte7}
    \Delta \longrightarrow \Delta_{F} =  F \, \Delta^{} \,
     F^{-1} \, ,\qquad S^{} \longrightarrow S_{F} = u \, S^{} \, u^{-1} \, ,
\end{equation}
where
\begin{eqnarray}\label{blte8}
%&\Delta^{(0)} (x) =  x \otimes 1 + 1 \otimes x,\quad S^{(0)}(x)=-x, \ \quad \forall x\in \mathfrak{g}
 %\cr\cr
&  F  = \sum\limits_{i} f^{(1)}_{i} \otimes  f^{(2)}_{i} \, , \qquad u =
\sum\limits_{i}  f^{(1)}_{i}\, S\, ( f^{(2)}_{i})\, .
\end{eqnarray}
If classical enveloping Lie algebra $U(\mathfrak{g})$ is considered as a Hopf algebra 
$H^{(0)} = (U (\mathfrak{g}), m, \Delta^{(0)}, S^{(0)}, \epsilon)$ then 
%after twisting procedure only the coalgebra sector (coproduct and coinverse) is modified. 
\begin{eqnarray}%\label{blte8}
&\Delta^{(0)} (x) =  x \otimes 1 + 1 \otimes x,\quad S^{(0)}(x)=-x, \ \quad \forall x\in \mathfrak{g}\,.\nonumber
 %\cr\cr
%&  F  = \sum\limits_{i} f^{(1)}_{i} \otimes  f^{(2)}_{i} \, , \qquad u =
%\sum\limits_{i}  f^{(1)}_{i}\, S\, ( f^{(2)}_{i})\, .
\end{eqnarray}

In order to get the coassociative coproduct one should postulate  the normalized 2-cocycle condition for the invertible twist element ${\mathcal F}$ (see \cite{Dr2})
\begin{equation}\label{coc}
F^{12}(\Delta \otimes id)\left( F\right) = F^{23}(id\otimes \Delta )\left(
F\right), \quad \left( \epsilon \otimes id\right)( F)=1=(id\otimes \epsilon) (F) .
\end{equation}%
In $H^{(0)}$ one can introduce the universal (quantum) $R$-matrix by the formula
\begin{equation}\label{Rtwist}
	R_F= F^\tau\, F^{-1}\,=\, (R_F^\tau)^{-1} \sim 1\otimes 1+r+o(\chi^2)
\end{equation}
which under the reality conditions (\ref{rc_1}) becomes unitary (at the same time real and antireal).
More generally, twist deformation of quasitriangular Hopf algebra $(H,R)$ give rise to quasitriangular Hopf algebra
with new universal $R$-matrix $R\longrightarrow R_F=F^\tau\,R\, F^{-1}$.
% provided that the formal parameter $\chi$ is assumed to be imaginary.

The simplest case one can deal  is an Abelian twist %$F_A=\exp{\xi A\wedge B}$, 
\begin{equation}\label{Atwist}
	F_{A,1/2}=\exp{(-{\chi\over 2}X\wedge Y)}, \qquad R_A=\exp{(\chi X\wedge Y)} 
\end{equation}
where two primitive commuting elements $[X,Y]=0$ determines $r_A=\chi X\wedge Y$ the skew-symmetric solutions of CYBE. 
In fact, the same Abelian quantum $R$-matrix $R_A$ can be implemented by the one-parameter family of Abelian twists
\begin{equation}\label{Ab}
	F_{A,s}=\exp{\xi(s\,X\otimes Y-(1-s)\,Y\otimes X)},\quad  s\in [0,1]
\end{equation}
 which are related with each other by a trivial (coboundary) twists . For example
\begin{equation}\label{Atwist2}
	F_{A,1}\equiv \exp{\chi\,X\otimes Y}= (W^{-1}\otimes W^{-1})\,F_{A,1/2}\,
	\Delta^{}(W),  \qquad  
\end{equation}
where $W=\exp({\xi\over 2}XY)$.
 \footnote{ We remind that a  coboundary twist for a given Hopf algebra is constructed out of any invertible element $W$ according to the following prescription 
 $$F_W^{cob}=(W^{-1}\otimes W^{-1})\Delta^{}(W)$$
 and leads via twisting (\ref{blte7}) to the isomorphic Hopf algebras (see e.g. \cite{Kulish09} ).}
%  corresponding to the similarity automorphism $X\longrightarrow W\,X\,W^{-1}$, for a suitable invertible element $W$.  This transformation leaves algebraic relations unchanged while changing coproduct within the isomorphism of Hopf algebra structure. 
 
Assuming  $X,Y$ real (antireal), i.e. $X^\divideontimes=\pm X, Y^\divideontimes=\pm Y$, the formal parameter $\chi$ has to be imaginary and all twist $F_{A,s}(\chi)$ are unitary. Consequently, anyone can be used to deform equivalently $\divideontimes$-Hopf algebras. However, there is an advantage of using (\ref{Atwist}). In this case the element $u$ (see  (\ref{blte8})) reduces to the unit and the antipodes map remains unchanged. %This property will be used several times later. 

If $X^\divideontimes=\pm Y, Y^\divideontimes=\pm X$ then only (\ref{Atwist})
is unitary for $\chi$ real. Thus the inverse transformation $F_{A,1/2}=(W^{}\otimes W^{})\,F_{A,1}\,\Delta^{}(W^{-1})$ can be treated as unitarizing the non-unitary twist $F_{A,1}$ by the coboundary twist $(W^{}\otimes W^{})\Delta^{}(W^{-1})$ (cf \ref{Atwist2}).

Alternatively, by introducing the (non-standard) flipped conjugation on the tensor product (see \cite{LNR91}) 
\begin{equation}\label{d-1f}
(a\otimes b)^\divideontimes = b^\divideontimes\otimes a^\divideontimes%%\qquad(\divideontimes-{\rm direct})~.
\end{equation}
and in the formulae (\ref{d-1}) one can regain all the twist $F_{A,s}$ unitary as well for the imaginary parameter $\chi$. This property will be used later  in
Sect. 5.4 for the case of quantized Abelian twist in a quantized Lorentz algebra.

\subsection{Two basic $\mathfrak{sl}(2; \mathbb{C})$ quantizations and their real versions}

\subsubsection{Quantization of Jordanian $r$-matrix $r_J$}

The quantum twist $F_{J_{}^{}}$ corresponding to the classical Jordanian
$r$-matrix $r_J$ is well known since a long time \cite{Og93} \footnote{Here $\chi$ is not an effective deformation
parameter. It is a formal variable which enables to write the twist as a formal power series and an invertible element. }
\begin{equation}\label{j1}
F_{J^{}}^{}(\chi)\,=\,\exp{(H\otimes\sigma}),\qquad\sigma\;=\;\ln(1+\chi E_{+})~.
\end{equation}
The twisted coproducts and antipodes are easy to derive %($k=0,1$)
\begin{eqnarray}\label{j2}
\Delta_{J}(E_{+})&=&{\mathcal F}(\chi)\Delta^{(0)}(E_+) {\mathcal F}^{-1}(\chi)=E_{+}\otimes e^{\sigma}+1\otimes E_{+}
\nonumber
\\
&& \nonumber\\ 
\Delta_{J}(H)
&= &H\otimes e^{-\sigma}+1\otimes H =H\otimes 1+1\otimes H-\chi H\otimes E_{+} e^{-\sigma}
%\nonumber \\[8pt]&&
%+\,(-1)^ki\beta\Sigma_{k+1}\otimes\Lambda_k e^{\Sigma_k}
%\nonumber
\\
&& \nonumber\\
\Delta_{J}(E_{-})&=&E_{-}\otimes e^{-\sigma}+1\otimes E_{-}+2\chi H\otimes H e^{-\sigma}
\nonumber
\\[8pt]
&&  -\chi^2 H (H-1)\otimes E_{+}e^{-2\sigma}\nonumber
 \end{eqnarray}
and
%Using above relations for the coproducts one calculates the following formulae for the antipodes
 \begin{eqnarray}\label{j3}
  S_{J}(E_{+})&=&-E_{+}\,e^{-\sigma}  ,\qquad\qquad  S_{J}(H)=-H\,e^{-\sigma}\nonumber\\[8pt]
  S_{J}(E_{-})&=&-E_{-}\,e^{\sigma}
  +2\chi H^2 e^{\sigma}
  +\chi^2 H (H-1)E_{+}e^{\sigma}%(e^{\Sigma_k}-1)
 % -2\chi^3\tau\,\sigma_{}^2E^2\ \ %(e^{\Sigma_k}-1)^2
  %\nonumber
%\\[8pt]
\end{eqnarray}
The quantum  $R$-matrix takes the form ($R_J=F_J^{21}F_J^{-1}$)
\begin{equation}\label{j4}
R_{J^{}}^{}(\chi)\,=\,F_J^{21}(\chi)F_J^{-1}(\chi )=\exp{(\sigma\otimes H)} \exp{(-H\otimes\sigma)} .
\end{equation}
The only compatible reality condition for the $\mathfrak{sl}(2;\mathbb{C})$ Jordanian deformation is of non-compact  $\mathsf{sl}_J(2)$ type (see \ref{rm6-4}) obtained if the parameter $\chi\in\imath\mathbb{R}$. In such case the Jordanian twist $F_J= \exp{(H\otimes\ln(1+\chi E_+))}$ is unitary, provides deformed coproducts and antipodes  satisfying automatically the conditions (\ref{d-1}). Therefore, it provides (see \ref{Rtwist}) the (real=antireal)  universal R-matrix $R_J=\exp{(\ln(1+\chi E_+)\otimes H)}\exp{(-H\otimes\ln(1+\chi E_+))}$.\smallskip\\
 
\subsubsection{Standard quantization of $\mathfrak{sl}(2;\mathbb{C})$}
%Another admissible quantum deformation of $U_{\mathfrak{sl}(2;\mathbb{C})}$ is   
The standard (non-triangular) quantum deformation is corresponding to the  solution of CYBE given by (\ref{rm5}). It is described  by $q$-analog or Drinfeld-Jimbo quantum deformation, with algebraic and coalgebraic sectors given by the following formulae
%\footnote{It can be shown by cohomological arguments that quantum generators $\E_{\pm}$ can be obtained from Lie algebra generators by non-linear transformation $\E_{\pm}\mapsto E_{\pm}+ o(\gamma)$ involving deformation parameter, i.e., that $U_{q}(\mathfrak{sl}(2;\mathbb{C}))\equiv U(\mathfrak{sl}(2;\mathbb{C}))[[\gamma]]$, where $q=\exp{{1\over 2}\gamma}$. However explicit form of such transformation is not known.}
\begin{eqnarray}\label{ex1}
q_{}^{\H}\E_{\pm}\!\!&=\!\!&q_{}^{\pm 1}\E_{\pm}\,q_{}^{\H}~,\quad q_{}^{\H_{}}q_{}^{-\H_{}}=q_{}^{-\H_{}}q_{}^{\H_{}}=1~,\quad
[\E_{+},\,\E_{-}]\;=\;\frac{q_{}^{2\H}-q_{}^{-2\H}}{q_{}^{}-q_{}^{-1}} %=\frac{\sinh(\gamma \H)}{\sinh({1\over 2}\gamma)}
~,
\\[7pt]\label{ex2}
\Delta_{q_{}}^{}(q_{}^{\pm \H_{}})\!\!&=\!\!&q_{}^{\pm \H_{}}\otimes
q_{}^{\pm \H_{}}~,  \qquad
\Delta_{q_{}}^{}(\E_{\pm})\;=\;\E_{\pm}\otimes
q_{}^{\H_{}}+q_{}^{-\H_{}}\otimes \E_{\pm}~,
\\[7pt]\label{ex3}
S_{q_{}}^{}(q_{}^{\pm \H_{}})\!\!&=\!\!&q_{}^{\mp \H_{}}~,\qquad
S_{q_{}}^{}(\E_{\pm})\;=\;-q_{}^{\pm 1}\E_{\pm}~,
\\[7pt]\label{ex4}
\epsilon_q(q_{}^{\pm\H})\!\!&=\!\!& 1~,\qquad \epsilon_q(\E_{\pm})=0~.
\end{eqnarray}
where we denote by $(q^\H, \E_\pm)$ the $q$-deformed or quantum CW basis. \footnote{Non-standard, e.g. Jordanian, deformation can be also expressed with the use of nonclassical quantum Lie algebra generators \cite{Delius95} obtained from twist (\ref{j1}) (see also \cite{ABP17}).}
The quantum universal $R$-matrix satisfying QYBE (\ref{i3b}) as well as the conditions (\ref{i2})-(\ref{i3a}) is given by the formula:
\begin{equation}\label{ex5b}
R_{q}=\exp_{q_{}^{-2}}\Bigl((q_{}^{}-q_{}^{-1})\E_{+}\,
q_{}^{-\H_{}}\otimes q_{}^{\H_{}}\E_{-}\Bigr) q_{}^{2\H_{}\otimes \H_{}}=
q_{}^{2\H_{}\otimes \H_{}}\exp_{q_{}^{-2}}\Bigl((q_{}^{}-q_{}^{-1})\E_{+}\,
q_{}^{\H_{}}\otimes q_{}^{-\H_{}}\E_{-}\Bigr) ~.
\end{equation}
where we use the standard definition of $q$-exponential $\exp_{q_{}^{-2}}$ (cf. Appendix B)
\begin{eqnarray}\label{ct16}
\exp_{q}(x)\!\!&:=\!\!&\sum_{n\geq0}\,\frac{x^n}{(n)_{q}^{}!}~,
\quad\;(n)_{q}^{}!:=(1)_{q}^{}(2)_{q}^{}\cdots
(n)_{q}^{},\quad(n)_{q}^{}=\frac{1-q^n}{1-q}~.
\end{eqnarray}
The alternative second version of the universal $R$-matrix has the form:
\begin{equation}\label{ex5b2}
R_{q}^{\tau-1}=\exp_{q_{}^{2}}\Bigl((q_{}^{-1}-q_{}^{})\E_{-}\,
q_{}^{-\H_{}}\otimes q_{}^{\H_{}}\E_{+}\Bigr) q_{}^{-2\H_{}\otimes \H_{}}=
q_{}^{-2\H_{}\otimes \H_{}}\exp_{q_{}^{2}}\Bigl((q_{}^{-1}-q_{}^{})\E_{-}\,
q_{}^{\H_{}}\otimes q_{}^{-\H_{}}\E_{+}\Bigr) ~.
\end{equation}
and provides nontrivial element $Q_q=R_qR_q^\tau$.
%\begin{equation}\label{ex5b3}
%Q_{q}=\exp_{q_{}^{-2}}\Bigl((q_{}^{}-q_{}^{-1})\E_{+}\,
%q_{}^{-\H_{}}\otimes q_{}^{\H_{}}\E_{-}\Bigr) q_{}^{2\H_{}\otimes \H_{}}=
%q_{}^{2\H_{}\otimes \H_{}}\exp_{q_{}^{-2}}\Bigl((q_{}^{}-q_{}^{-1})\E_{+}\,
%q_{}^{\H_{}}\otimes q_{}^{-\H_{}}\E_{-}\Bigr) ~.
%\end{equation}
This quantum $R$-matrices describe by their linear terms, in the limit $\gamma\mapsto 0, q\mapsto 1$,  non-skewsymmetric  classical $r$-matrices in the Belavin-Drinfeld form (\ref{rm5}).
 
Three standard real  forms (\ref{rm6-1})-(\ref{rm6-3}) impose the following reality conditions on $q$-deformed   
generators $(q^{\pm\H},\E_\pm)$:
 \begin{eqnarray}\label{ex6-1}                                                                                      (q^\H)^{\dagger}\!\!&=\!\!&q^\H,\qquad \E_{\pm}^{\dagger}\;=\;\E_{\mp},\quad q\in\mathbb{R}\Leftrightarrow\gamma\in \mathbb{R}                        \quad{\rm{for}}\;\;\mathsf{su}_\gamma(2),\\[5pt]\label{ex6-2}
(q^\H)^{\#}\!\!&=\!\!&q^\H,\quad\;\E_{\pm}^{\#}\;=\;-\E_{\mp},\quad\;  q\in\mathbb{R}\Leftrightarrow\gamma\in \mathbb{R}                  \quad\;{\rm{for}}\;\;\mathsf{su}_\gamma(1,1), \\[5pt]\label{ex6-3}                                            (q^\H)^{\star}\!\!&=\!\!&q^\H,\quad\;\E_{\pm}^{\star}\;=\;-\E_{\pm},\quad\;\; |q| =1\Leftrightarrow\gamma\in \imath\mathbb{R}            \quad{\rm{for}}\;\;\mathsf{sl}_\gamma(2), 
\end{eqnarray}
which turn, in each case, the Hopf algebra (\ref{ex1})--(\ref{ex4}) into the real Hopf algebra satisfying the reality conditions (\ref{d-1}).
Taking into consideration the restriction on the values of $q$ one can see that reality conditions
(\ref{rm6-1})-(\ref{rm6-3}) for $(H, E_\pm)$ have the same form  as for $(\H, \E_\pm)$ (see (\ref{ex6-1})-(\ref{ex6-3})).
The last two (non-compact) real forms coincide in the classical limit $\gamma\mapsto 0$.  In the classical limit $\gamma\mapsto 0$ deformed and undeformed generators can be identified, i.e. $\H\mapsto H,\ \E_\pm\mapsto E_\pm,\ q^\H\mapsto 1$. For the  first two real forms the corresponding universal R-matrix (\ref{ex4}) is real,  
the last case (\ref{rm6-3}) is antireal.

We recall that however if the Jordanian Lie bialgebra has no effective deformation parameter its quantization requires the introduction of  a (formal) parameter $\chi$, which permits to construct the twist and the quantum $R$-matrix as a formal power series, elements of $U(\mathfrak{sl}(2; \mathbb{C}))\otimes U(\mathfrak{sl}(2; \mathbb{C}))[[\chi]]$. In contrast,  Lie bialgebras corresponding to standard deformations are parametrized by numerical (complex or real) factor $\gamma$, describing effective deformation parameter. 
%They can be quantized using two different approaches:
%
%i) the parameter $\gamma$ is formal, thus the element $q^\H=\exp{{1\over 2}\gamma \H}\in U(\mathfrak{sl}(2; \mathbb{C}))[[\gamma]]$   becomes a formal power series;
%
%ii) one introduces new generators $\texttt{k}^{\pm 1}=q^{\pm \H}$ and rewrites Hopf algebra defining relations 
%(\ref{ex1}) -- (\ref{ex1}) in terms of $(\texttt{k}^{\pm 1}, \E_{\pm})$ generators with the numerical  value
 %$q=\exp{{1\over 2}\gamma}$ (so-called specialization), where $\gamma\in\mathbb{C}$ parametrizes  corresponding Lie bialgebra structure $\mathsf{su}_\gamma(2)$, $\mathsf{su}_\gamma(1, 1)$, or $\mathsf{sl}_\gamma(2)$. 
%%The case when $q$ is $N$-th root of unity, i.e. $q=\exp(\imath {2\pi\over N})$, should be treated separately.
%In this paper, in order to be compatible with Jordanian deformation,  we shall utilize  $h$-addic point of view.

\setcounter{equation}{0}
%\section{Lie algebra of complex $D=4$ rotations $\mathfrak{o}(4;\mathbb{C})\cong\mathfrak{sl}(2;\mathbb{C})\oplus \mathfrak{sl}(2;\mathbb{C})$ and its real forms}
\section{Lie bialgebras of complex $D=4$ rotations and their  real forms}

In this section we describe Lie bialgebra of $D=4$ complex rotations $\mathfrak{o}(4;\mathbb{C})$ \footnote{It is known that complex metric (symmetric, nondegenerate and bilinear form) has no signature}  and its real forms: Euclidian, Lorentz, Kleinian and quaternionic orthogonal Lie algebras
%\footnote{It should be noted that the names ''Euclidean'' and Kleinian are used also for denomination of inhomogeneous Lie symmetries: rotations with translations generated by fourmomenta.} 
in terms of chiral left $(H,E_\pm)$ and right $(\bar H, \bar E_\pm)$ CW bases:   %$\mathfrak{sl}(2;\mathbb{C})\oplus\bar{\mathfrak{sl}}(2;\mathbb{C})$ basis,
\footnote{For the relation with other, physically more meaningful, Cartesian basis see e.g. (\ref{jl1}) and \cite{BLTnov15}.}
\begin{eqnarray}\label{3o}
[H,\,E_{\pm}]\!\!&=\!\!&E_{\pm}~,\quad[E_{+},\,E_{-}]\,=\,2H~,\quad[\bar{H},\,\bar{E}_{\pm}]\,=\,\bar{E}_{\pm}~,\quad [\bar{E}_{+},\,\bar{E}_{-}]\,=\,2\bar{H}~.
\end{eqnarray}
%It appears that for description of quantum deformations (cf. Subsect, 2.2) and in particular for the classification of classical $r$-matrices %of the  complex Euclidean algebra $\mathfrak{o}(4;\mathbb{C})$
%and its real forms it is convenient to use the Cartan--Weyl bases in both sectors of the sum $\mathfrak{o}(4;\mathbb{C})=\mathfrak{sl}(2;\mathbb{C})\oplus\bar{\mathfrak{sl}}(2;\mathbb{C})$. 
Due to the fact that each $\mathfrak{sl}(2;\mathbb{C})$ sector has two bialgebra structures (single Jordanian and standard one-parameter family) one can easily to identify three (up to the flip) types of bialgebra structures on $\mathfrak{o}(4;\mathbb{C})$, namely the direct sums
\begin{eqnarray}\label{3o1}
\mathsf{o}_{\gamma,\bar\gamma}(4;\mathbb{C})\!\!&=\!\!&\mathsf{sl}_\gamma(2;\mathbb{C})\oplus\bar{\mathsf{sl}}_{\bar\gamma}(2;\mathbb{C}),
\\ \label{3o2}
\mathsf{o}_{\gamma,\bar J}(4;\mathbb{C})\!\!&=\!\!&\mathsf{sl}_\gamma(2;\mathbb{C})\oplus\bar{\mathsf{sl}}_{J}(2;\mathbb{C}).
\\ \label{3o3}
\mathsf{o}_{J,\bar J}(4;\mathbb{C})\!\!&=\!\!&\mathsf{sl}_J(2;\mathbb{C})\oplus\bar{\mathsf{sl}}_J(2;\mathbb{C}).
\end{eqnarray}
with the classical $r$-matrices obtained by summing up the pair of  chiral and antichiral contributions, e.g. $r_{J,\bar J}=r_J+\bar r_J=H\wedge E_+
+\bar H\wedge \bar E_+$ in (\ref{3o3}), etc.
The list (\ref{3o1})--(\ref{3o3}) does not exhaust all possible bialgebra structures because it does not take into account the mixed terms belonging to $\mathfrak{sl}(2;\mathbb{C})\wedge\bar{\mathfrak{sl}}(2;\mathbb{C})$, which 
can also contribute to the classical $r$-matrices.    
In \cite{BLTnov15} using purely algebraic methods we classified all $\mathfrak{o}(4;\mathbb{C})$ bialgebras. 
We found five families of complex skewsymmetric $r$-matrices:  three, each with three-parameters, one two-parameter and one with one parameter.% $r$-matrices  which 
The list of $\mathfrak{o}(4;\mathbb{C})$ $r$-matrices looks as follows \cite{BLTnov15}:
\footnote{The list (\ref{r1}) -(\ref{r5}) is numbered in different way in comparison with
original result \cite{BLTnov15}; the $r$-matrix $r_6$ in \cite{BLTnov15} from $r_5=r_V$ by involutive automorphism
flipping the chiral sectors. Notation for the parameters is slightly changed as well.}
\begin{eqnarray}
%\begin{array}{rcl}
r_{I}^{}(\chi)\!\!&=\!\!&\chi(E_{+}+\bar{E}_{+})\wedge(H+\bar{H})~,
\label{r1} \\[7pt]
r_{II}^{}(\chi,\bar{\chi},\varsigma)\!\!&=\!\!&\chi\,E_{+}\wedge H+\bar{\chi}\,\bar{E}_{+}\wedge\bar{H}+\varsigma E_{+}\wedge\bar{E}_{+}~,
\label{r2}\\[7pt]
r_{III}^{}(\gamma,\bar{\gamma},\eta)\!\!&=\!\!&\gamma\,E_{+}\wedge E_{-} +\bar{\gamma}\,\bar{E}_{+}\wedge\bar{E}_{-} + \eta\,H\wedge\bar{H}~,
\label{r3}\\[7pt]
r_{IV}^{}(\gamma,\varsigma)\!\!&=\!\!&\gamma\left(E_{+}\wedge E_{-} -\bar{E}_{+}\wedge\bar{E}_{-} -2H\wedge\bar{H}\right)+\varsigma E_{+}\wedge\bar{E}_{+}
\label{r4}\\[7pt]
r_{V}^{}(\gamma,\bar{\chi},\rho)\!\!&=\!\!&\gamma\,E_{+}\wedge E_{-} +\bar{\chi}\,\bar{E}_{+}\wedge\bar{H}+\rho H\wedge\bar{E}_{+}~.   \label{r5}
%\\[7pt]\label{r6}
%r_{6}^{}(\bar{\gamma},\chi,\bar\rho)\!\!&=\!\!&\bar{\gamma}\,\left( \bar{E}_{+}\otimes\bar{E}_{-}+\bar H\otimes\bar H\right)+\chi\,E_{+}\wedge H+\bar \rho {E}_{+}\wedge\bar{H}~.
%\end{array}
\end{eqnarray}
Here all parameters $\gamma$, $\bar{\gamma}$, $\eta$, $\chi$, $\bar{\chi}$, $\varsigma$, $\rho$, $\bar{\rho}$  are
arbitrary complex numbers and they are independent in different $r$-matrices.

The first two $r$-matrices $r_{I}^{}(\chi)$ and $r_{II}^{}(\chi,\bar{\chi},\varsigma)$, generate twist and they satisfy the homogeneous CYBE (\ref{i4}).
%with $\Omega=0$, i.e. when $\gamma,\bar{\gamma}=0$ in (\ref{qdc3}).
Moreover the first $r$-matrix $r_{I}^{}(\chi)$ is pure Jordanian type and the second $r$-matrix $r_{II}^{}(\chi,\bar{\chi},\varsigma)$ is the sum of two Jordanian ones with
third one describing Abelian twist: $r_{II}^{}(\chi,\bar{\chi},\varsigma)= r_{II}^{}(\chi,0,0)+r_{II}^{}(0,\bar{\chi},0)+r_{II}^{}(0,0,\varsigma)$.
The third $r$-matrix $r_{III}^{}(\gamma,\bar{\gamma},\eta)$ is the sum of two standard $r$-matrices and one Abelian: $r_{III}^{}(\gamma,\bar{\gamma},\eta)= r_{III}^{}(\gamma,0,0)+r_{III}^{}(0,\bar{\gamma},0)+r_{III}^{}(0,0,\eta)$.  The fourth $r$-matrix $r_{IV}^{}(\gamma,\chi')$ is the sum of special choice of the third $r$-matrix and the Abelian  $r$-matrix: $r_{IV}^{}(\gamma,\varsigma):=r_{III}^{}(\gamma,-\gamma,-2\gamma)+\varsigma E_{+}\wedge\bar{E}_{+}$.
The last $r$-matrices $r_{V}^{}(\gamma,\bar{\chi},\rho)$  is the sum of standard, Jordanian and  Abelian $r$-matrices: $r_{V}^{}(\gamma,\bar{\chi},\rho)=r_{V}^{}(\gamma,0,0)+r_{V}^{}(0,\bar{\chi},0)+r_{V}^{}(0,0,\rho)$.
The formulae  for ($r_{II}, r_{III}, r_{V}$) are obtained by supplementing (\ref{3o1}) - (\ref{3o3}) with particular additional Abelian contributions belonging to $\mathfrak{sl}(2;\mathbb{C})\wedge\bar{\mathfrak{sl}}(2;\mathbb{C})$. % depending on Cartan generators $H, \bar H$.

%We recall that non-standard (twist) deformation of $\mathfrak{sl}(2;\mathbb{C})$ is fully described by its skew-symmetric classical $r$-matrices $r_J=H\wedge E_+$ satisfying CYBE. In contrast, the antisymmetric $r$-matrix $E_+\wedge E_-$ describing the standard (quasi-triangular)  deformation satisfies mCYBE. Its non-symmetric counterpart satisfying CYBE is $E_{+}\otimes E_{-}+ H\otimes H$.  This form is necessary in order to distinguish between real and anti-real cases after imposing reality conditions. Therefore one should rewrite matrices $r_{III}, r_{IV}, r_V$ in the following form
We shall calculate as well in next Section for all five quantizations generated by (\ref{r1})--(\ref{r5}) the universal $R$-matrices. Using formula (\ref{i1}) one obtains in third, fourth and fifth cases the following Belavin-Drinfeld type of matrices which appear in the expansion (\ref{i1}): 
\begin{eqnarray}
\begin{array}{rcl}
 \tilde r_{III}^{}(\gamma,\bar{\gamma},\eta)\!\!&=\!\!&\gamma\,\left(E_{+}\otimes E_{-}+ H\otimes H \right)+\bar{\gamma}\,\left( \bar{E}_{+}\otimes\bar{E}_{-}+\bar H\otimes\bar H\right)+\eta\,H\wedge\bar{H}~,
\label{r3t}\\[7pt]
\tilde r_{IV}^{}(\gamma,\varsigma)\!\!&=\!\!&\gamma\,\left(E_{+}\otimes E_{-}+H\otimes H -\bar{E}_{+}\otimes\bar{E}_{-}-\bar H\otimes\bar H -2H\wedge\bar{H}\right)+\varsigma E_{+}\wedge\bar{E}_{+}
\label{r4t}\\[7pt]
\tilde r_{V}^{}(\gamma,\bar{\chi},\rho)\!\!&=\!\!&\gamma\,\left(E_{+}\otimes E_{-}+ H\otimes H \right)+\bar{\chi}\,\bar{E}_{+}\wedge\bar{H}+\rho H\wedge\bar{E}_{+}~.   \label{r5t}
%\\[7pt]\label{r6}
%r_{6}^{}(\bar{\gamma},\chi,\bar\rho)\!\!&=\!\!&\bar{\gamma}\,\left( \bar{E}_{+}\otimes\bar{E}_{-}+\bar H\otimes\bar H\right)+\chi\,E_{+}\wedge H+\bar \rho {E}_{+}\wedge\bar{H}~.
\end{array}
\end{eqnarray}
%It is well-known from the theory of real forms for semisimple complex Lie algebras (including our case $\mathfrak{o}(4;\mathbb{C})$) that all real (non-compact) forms can be constructed by involutive automorphisms of 
There is unique compact real form $\mathfrak{o}(4)$ and %(see \cite{Gant39})  and 
%For each such Lie algebra the compact form is unique (see e.g. \cite{Gant39}). Besides there are only 
three real non-compact forms of $\mathfrak{o}(4)$: %(see e.g. \cite{BarutRacz77}):
{\it the Lorentz algebra $\mathfrak{o}(3,1):=\mathfrak{o}(3,1;\mathbb{R})\cong\mathfrak{sl}(2;\mathbb{C})^{\mathbb{R}}$}, {\it the Kleinian algebra $\mathfrak{o}(2,2):=\mathfrak{o}(2,2;\mathbb{R})\cong\mathfrak{o}(2,1)\oplus\mathfrak{o}(2,1)$} and {\it the quaternionic Lie algebra $\mathfrak{o}^{\star}(4):=\mathfrak{o}(2;\mathbb{H})\cong\mathfrak{o}(2,1)\oplus\mathfrak{o}(3)$}.
%The algebra $\mathfrak{o}(4;\mathbb{C})$ has four real forms where
These  real forms 
%(the compact Euclidean algebra $\mathfrak{o}(4)$, the noncompact quaternionic symmetry $\mathfrak{o}^{\star}(4)$ and the noncompact Kleinian algebra $\mathfrak{o}(2,2)$) %preserve the chiral decomposition (\ref{rf1}), i.e. they
can be  expressed as the following six direct sums %(up to the flip automorphism) 
of $\mathfrak{sl}(2;\mathbb{C})$- real forms listed in (\ref{rm6-1})--(\ref{rm6-3}) %$\mathfrak{o}(3)$ and $\mathfrak{o}(2,1)$:
\begin{eqnarray}\label{rf1}
&H^{\dagger}=H,\quad E_{\pm}^{\dagger}=E_{\mp},\quad\bar{H}^{\dagger}=\bar{H},\quad\bar{E}_{\pm}^{\dagger}=\bar{E}_{\mp}\quad{\rm{for}}\;\mathfrak{o}(4),
\\[5pt]
&\begin{array}{l}\label{rf2}
{H}^{\dagger}=H,\quad E_{\pm}^{\dagger}=E_{\mp},\quad(\bar{H})^{\#}=\bar{H},\quad(\bar{E}_{\pm})^{\#}=-\bar{E}_{\mp},
\\[2pt]
H^{\dag}=H,\quad E_{\pm}^{\dag}=E_{\mp},\quad(\bar{H})^{\star}=-\bar{H},\quad(\bar{E}_{\pm})^{\star}=-\bar{E}_{\pm}
\end{array}\;\;{\rm{for}}\;\mathfrak{o}^{\star}(4),
\\[5pt]
&\begin{array}{l}\label{rf3}
{H}^{\#}=H,\quad {E_{\pm}}^{\#}=-E_{\mp},\quad(\bar{H})^{\#}=\bar{H},\quad(\bar{E}_{\pm})^{\#}=-\bar{E}_{\mp},
\\[2pt]
{H}^{\#}=H,\quad {E_{\pm}}^{\#}=-E_{\mp},\quad(\bar{H})^{\star}=-\bar{H},\quad(\bar{E}_{\pm})^{\star}=-\bar{E}_{\pm},
\\[2pt]
{H}^{\star}=-H,\quad{E_{\pm}}^{\star}=-E_{\pm},\quad(\bar{H})^{\star}=-\bar{H},\quad(\bar{E}_{\pm})^{\star}=-\bar{E}_{\pm}
\end{array}{\rm{for}}\;\mathfrak{o}(2,2),
\\[5pt]\label{rf4}
&H^{\ddag}=-\bar{H},\quad E_{\pm}^{\ddag}=-\bar{E}_{\pm},\quad
(\bar{H}^{\ddag}=-H,\quad\bar{E}_{\pm}^{\ddag}=-E_{\pm})\quad{\rm{for}}\;\mathfrak{o}(3,1).
\end{eqnarray}
The last real form (\ref{rf4}) characterizing the Lorentz  $\mathfrak{o}(3,1)$-algebra, does not preserve the chiral decomposition.

By imposing all real involutions in the list of classical  complex $r$-matrices (\ref{r1})--(\ref{r5}) we get complete set of real bialgebra structures on  the Lie algebra $\mathfrak{o}(4;\mathbb{C})$. 
The list of all real bialgebras for $\mathfrak{o}(4;\mathbb{C})$, together with  specified values for the corresponding  parameters, is presented in the table below, where $\mathfrak{o}^\star(4)$, $\mathfrak{o'}^{\star}(4)$ denotes the bialgebras after imposing the reality conditions (\ref{rf2}), and  $\mathfrak{o}''(2,2)$, $\mathfrak{o}'(2,2)$, $\mathfrak{o}''(2,2)$
denotes three bialgebras obtained by applying three reality conditions (\ref{rf3}).
%\footnote{When quantized these parameters become formal. Thus e.g. $\gamma\in\mathbb{R}\Longleftrightarrow (\gamma)^\star=\gamma$.}.
%%\newline
\begin{center}
% {\bf Table A. All real Lie bialgebra structures for $\mathfrak{o}(4;\mathbb{C})$ }\\[1ex]
\begin{table}[h]
\begin{tabular}[h]{|c|c|c|c|c|c|}
\hline
&&&&\\[-11pt]
• &$ r_I(\chi) $&$ r_{II}(\chi, \bar\chi,\varsigma)$& $ r_{III}(\gamma, \bar\gamma,\eta)$&$ r_{IV}(\gamma, \varsigma)$&$ r_V(\gamma, \bar\chi,\rho)$\\
\hline
%• & $\chi$& $(\chi, \bar\chi,\varsigma)$& $(\gamma, \bar\gamma,\eta)$& $(\gamma, \varsigma)$ & $(\gamma, \bar\chi,\rho)$\\
%\hline
&&&&\\[-11pt]
$\mathfrak{o}(4)$ & • & • & $\gamma,\bar\gamma\in\mathbb{R}$ ; $\eta\in\imath\mathbb{R}$&  •   & • \\
\hline
&&&&\\[-11pt]
$\mathfrak{o}^\star(4)$ & • & • &  $\gamma,\bar\gamma\in\mathbb{R}$ ; $\eta\in\imath\mathbb{R}$& • & • \\
\hline
&&&&\\[-11pt]
 $\mathfrak{o'}^{\star}(4)$ & • & • &  $\gamma,\eta\in\mathbb{R}$ ; $\bar\gamma\in\imath\mathbb{R}$  & • & $\gamma,\rho\in\mathbb{R}$ ; $\bar\chi\in\imath\mathbb{R}$   \\
\hline
&&&&\\[-11pt]
$\mathfrak{o}(2,2)$ & • & • &  $\gamma,\bar\gamma\in\mathbb{R}$ ; $\eta\in\imath\mathbb{R}$  & • & • \\
\hline
&&&&\\[-11pt]
$\mathfrak{o}'(2,2)$ & • & • &  $\gamma,\eta\in\mathbb{R}$ ; $\bar\gamma\in\imath\mathbb{R}$ & • & $\gamma,\rho\in\mathbb{R}$ ; $\bar\chi\in\imath\mathbb{R}$  \\
\hline
&&&&\\[-11pt]
$\mathfrak{o}''(2,2)$ &  $\chi\in\imath\mathbb{R}$  &  $\chi,\bar\chi, \varsigma\in\imath\mathbb{R}$   &  $\gamma,\bar\gamma, \eta\in\imath\mathbb{R}$   & $\gamma,\varsigma\in\imath\mathbb{R}$  &  $\gamma,\bar\chi, \rho\in\imath\mathbb{R}$    \\
\hline
&&&&\\[-11pt]
$\mathfrak{o}(3,1)$& $\chi\in\imath\mathbb{R}$ &  $\chi=\bar\chi\in\imath\mathbb{R}$ ; $\varsigma\in\mathbb{R}$ &  $\bar\gamma=-\gamma^*\in\mathbb{C}$ ; $\eta\in\mathbb{R}$  & $\gamma,\varsigma\in\mathbb{R}$ & • \\
\hline
%&&&&\\[-10pt]
\end{tabular}
\caption{All real Lie bialgebras for $\mathfrak{o}(4;\mathbb{C})$}
	\label{tab:Real}
\end{table}
\end{center}
%\newpage
%%%%%%% 

In the following Section we shall describe the Hopf-algebraic quantization of five complex $\mathfrak{o}(4;\mathbb{C})$ $r$-matrices (\ref{r1})--(\ref{r5}).
%The twist quantization of $r_I, r_{II}$ does not imply the change of algebra generators and one can use the reality
%conditions expressed in classical basis. For $r$-matrices $r_{III}, r_{IV}, r_V$ containing Drinfeld-Jimbo $q$-analogues
%of classical algebra basis  we shall express the reality conditions   using $q$-deformed generators (see
%(\ref{ex1})-- (\ref{ex3}) ) which however for presented restrictions on $q$-parameter %(see also (\ref{ex6-1})--(\ref{ex6-3})) 
%have the same form for $q$-analogs $(\H, \E_\pm)$ as for the classical generators $(H, E_\pm)$.
Out of these five complex quantizations after imposing seven reality conditions we obtain sixteen real 
$\mathfrak{o}(4;\mathbb{C})$ Hopf algebra structures: $r_{III}$ provides seven real forms,  $r_V$ -- three, and each of remaining three leads to two real quantizations.

\setcounter{equation}{0}
\section{Explicit quantizations of $\mathfrak{o}(4;\mathbb{C})$ and their real forms}

\subsection{Jordanian quantization of $\mathfrak{o}(4;\mathbb{C})$ ($r$-matrix $r_I$)}

Following the previous considerations (Subsect. 2.2) the quantum twist $F_{1_{}^{}}$ corresponding to the classical Jordanian $r$-matrix (\ref{r1}) can be written as
\begin{equation}\label{J1}
F_{1^{}}^{}(\chi)\,=\,\exp{((H+\bar H)\otimes\sigma}),\qquad\sigma\;=\;\ln(1+\chi (E_{+}+\bar E_+))~,
\end{equation}
Coproducts and antipodes are easy to derive (cf. (\ref{j1})--(\ref{j4})) %($k=0,1$)
\begin{eqnarray}\label{J2}%blte21
\Delta_{1}(E_{k+})
&=&{\mathcal
F}(\chi)\Delta^{(0)}(E_k)
{\mathcal F}^{-1}(\chi)=\Delta_1(E_{k+})=E_{k+}\otimes e^{\sigma}+1\otimes E_{k+}
\nonumber
\\
&& \nonumber\\
\Delta_{1}(H_k)
&= & H _k\otimes 1+1\otimes H_k-\chi(H+\bar H)\otimes E_{k+} e^{-\sigma}
%\nonumber \\[8pt]&&
%+\,(-1)^ki\beta\Sigma_{k+1}\otimes\Lambda_k e^{\Sigma_k}
\nonumber\\
&& \\
\Delta_{1}(E_{k-})&=&E_{k-}\otimes e^{-\sigma}+1\otimes E_{k-}
+2\chi (H+\bar H)\otimes H_ke^{-\sigma}
\nonumber
\\[8pt]
&&
  -\chi^2 (H+\bar H)(H+\bar H-1)\otimes E_{k+}e^{-2\sigma}\nonumber
 \end{eqnarray}
%%%%%%%%%%
where $k\in \{0, 1\}\equiv \mathbb{Z}_2$ and in order to reduce the  number of formulae we denoted  
$X_0= \{H=H_0, E_\pm=E_{0\pm}\}$
and $X_1= \{\bar H=H_1, \bar E_\pm=E_{1\pm}\}$. \footnote{The same convention will be further used below in the paper.}

Similarly, the formulae for the antipodes look as follows
 \begin{eqnarray}\label{J3}%{blte22a}
  S_{1}(E_{k+})&=&-E_{k+}\,e^{-\sigma}  ,\qquad\qquad  S_{1}(H_k)=-H_k -\chi (H+\bar H) E_{k+}\nonumber\\[8pt]
  S_{1}(E_{k-})&=&-E_{k-}\,e^{\sigma}
  +2\chi (H+\bar H)H_k e^{\sigma}
  +\chi^2 (H+\bar H)(H+\bar H-1)E_{k+}e^{\sigma}%(e^{\Sigma_k}-1)
 % -2\chi_k^3\tau\,\Sigma_{k+1}^2E_k^2\ \ %(e^{\Sigma_k}-1)^2
  %\nonumber
%\\[8pt]
\end{eqnarray}
The universal quantum  $R$-matrix takes the form ($R=F^{21}F^{-1}$)
\begin{equation}\label{J4}
R_{1^{}}^{}(\chi)\,=\,\exp{(\sigma\otimes (H+\bar H)}) \exp{(-(H+\bar H)\otimes\sigma}) .
\end{equation}

This simple one-parameter deformation admits two real quantum group structures (cf. Table \ref{tab:rr1}) as
indicated below. Since the twist is Jordanian, the reality conditions (\ref{d-1}) are valid if the 
%is automatic since deformation is performed by the unitary twist provided 
deformation parameter $\chi$ is imaginary. The Lorentzian case requiring as well imaginary $\chi$ has been already  studied in \cite{BLT08} with more details.

\begin{center}
% {\bf Table A. All real Lie bialgebra structures for $\mathfrak{o}(4;\mathbb{C})$ }\\[1ex]
\begin{table}[h]
\begin{tabular}{|c|c|c|c|}
\hline
&&&\\[-8pt]
$\mathfrak{o}''(2,2)$ &    $\ \ \ \ \ \chi\in\imath\mathbb{R}\ \ \ \ \ $   &
$H^{\star}=-H, E_{\pm}^{\star}=-E_{\pm}$
& $\bar H^{\star}= -\bar H , \bar E_{\pm}^{\star}=-\bar E_{\pm}$\\
\hline
&&&\\[-8pt]
$\mathfrak{o}(3,1)$  &  $\ \ \ \ \ \chi\in\imath\mathbb{R}\ \ \ \ \ $   & $H^{\ddagger} =-\bar{H} , E_{\pm}^{\ddagger}=-\bar E_{\pm}$
& $\bar H^{\ddagger}= -H  , \bar E_{\pm}^{\ddagger}=- E_{\pm}$\\
\hline
\end{tabular}
\caption{Real quantizations of $r_{I}(\chi)=\chi(E_{+}+\bar{E}_{+})\wedge(H+\bar{H})$}
	\label{tab:rr1}
\end{table}
\end{center}

\subsection{Left and right Jordanian quantizations intertwined by Abelian twist ($r$-matrix $r_{II}$) }

We see that for $\varsigma=0$ the $r$-matrix (\ref{r2}) describes two complex Jordanian $r$-matrices, each one for  chiral sectors $\mathfrak{sl}(2,\mathbb{C})$ and $\overline{\mathfrak{sl}(2;\mathbb{C})}$. They do commute with each other and can be quantized as the product of two Ogievetsky twists ($k=1,2$) (\cite{Og93} see also (\ref{j1}))
\begin{equation}\label{e10}
    F_{J,0}(\chi) = \exp{(H\otimes \Sigma)} \,\quad  F_{J,1}(\bar\chi) = \exp{(\bar H\otimes \bar\Sigma)}
\end{equation}
where $\Sigma=ln{(1+\chi E_+)}$, $\bar\Sigma=ln{(1+\bar\chi \bar E_+)}$. 
%So at this stage the quantized enveloping algebra splits into a tensor product of two Jordanian Hopf algebras (cf. ()) with independent deformation parameters. 
The next step is to consider the Abelian part of the classical $r$-matrix $r_{II}$ belonging to
$\mathfrak{sl}(2;\mathbb{C})\wedge\bar{\mathfrak{sl}}(2;\mathbb{C})$ intertwining two chiral coalgebra sectors
which ceases to be independent. % making them {it interacting} each other.
Because the generators $(H,\,E_{\pm})$ and $(\bar H,\, \bar E_\pm)$ do commute the twist function corresponding to (\ref{r2}) is given by the following formula:
 \begin{equation}\label{e11}
     F_2(\chi,\bar\chi, \varsigma)=  F_A (\chi,\bar\chi, \varsigma) F_{J,1}(\bar\chi) F_{J,0}(\chi) =
     F_A (\chi,\bar\chi, \varsigma)  F_{J,0}(\chi) F_{J,1}(\bar\chi)\,.
\end{equation}
where the Abelian twist $F_A$ takes the form\footnote{The normalization $\frac{\varsigma}{\chi\bar\chi}$ in the deformation parameter is necessary in order to
recover correct formula in the limit $\chi, \bar\chi\mapsto 0$.}% i.e. when purely Abelian twist comes into play.}
\begin{equation}\label{e12}
F_A(\chi,\bar\chi, \varsigma)=\exp{(\frac{\varsigma}{\chi\bar\chi}\Sigma\wedge\bar\Sigma)}
\end{equation}
which follows from the property  that elements $\Sigma,\ \bar\Sigma$ are primitive after performing Jordanian deformation.
We would like to mention here that the form of the twist function
given above  by formula (\ref{e11}) was proposed firstly %with not antisymmetrized exponential factor in (\ref{e12})
by Kulish and Mudrov  \cite{KM99}. %\newpage
% who listed as well all quantum deformations of the $D=4$ Lorentz algebra.
If we use (\ref{blte7}--\ref{blte8}),  and (\ref{e11}) we obtain the following formulae for the coproducts of $sl(2;\mathbb{C})\oplus \bar{sl}(2;\mathbb{C})$ generators  $(H_k,\, E_{k+},\,E_{k-})$, $k=0,1\in \mathbb{Z}_2$
 \begin{eqnarray}\label{e21}
\Delta_{2}(E_{k+})
&=&{\mathcal
F}(\chi,\bar\chi,\varsigma)\Delta^{(0)}(E_k)
{\mathcal F}^{-1}(\chi,\bar\chi,\varsigma)=E_{k+}\otimes e^{\Sigma_k}+1\otimes E_{k+}\nonumber
\\\nonumber
&& \\\nonumber
\Delta_{2}(H_k)
&= & H _k\otimes e^{-\Sigma_k}+1\otimes H_k\, + %-(-1)^k \frac{\chi^\prime}{\chi_{k+1}}\, E_{k+} e^{-\Sigma_k}\otimes\Sigma_{k+1} e^{-\Sigma_k}
\nonumber \\[8pt]&&
(-1)^k \frac{\varsigma}{\chi_{k+1}}\,\left(\Sigma_{k+1}\otimes E_{k+} e^{-\Sigma_k}-\, E_{k+} e^{-\Sigma_k}\otimes\Sigma_{k+1} e^{-\Sigma_k}\right)\nonumber
\\\nonumber
&& \\%\nonumber
\Delta_{2}(E_{k-})&=&E_{k-}\otimes e^{-\Sigma_k}+1\otimes E_{k-}
+2\chi_k H_k\otimes H_ke^{-\Sigma_k}
 +\chi_k H_k(H_k-1)\otimes\Lambda_k+%\nonumber 
\\[8pt]&&
\,(-)^k\frac{2\varsigma}{\chi_{k+1}}\,\left( H_ke^{-\Sigma_k}\otimes\Sigma_{k+1}e^{-\Sigma_k} - H_k\Sigma_{k+1}\otimes\Lambda_k
-\Sigma_{k+1}\otimes H_ke^{-\Sigma_k}\right)
\nonumber \\[8pt]&&
\,(-)^k\frac{2\varsigma}{\chi_{k+1}}\, \left(\,
\Lambda_k e^{\Sigma_k}\otimes H_k\Sigma_{k+1}e^{-\Sigma_k}\,+
H_k\Lambda_ke^{\Sigma_k}\otimes\Sigma_{k+1}\Lambda_k\,
%+\Sigma_{k+1}\otimes\Lambda_k\,-\Lambda_k\otimes\Sigma_{k+1}e^{-\Sigma_k}
\right)
\nonumber \\[8pt]&&
 +\,(-)^k\,\frac{\varsigma}{\chi_{k+1}}\,\left(
 %2H_k\Lambda_ke^{\Sigma_k} +%-2\,\Lambda_k \Sigma_{k+1} e^{\Sigma_k}
 \left(1-e^{-2\Sigma_k}\right)\otimes\Sigma_{k+1}\Lambda_k\, +\Sigma_{k+1}\otimes\Lambda_k\,-\Lambda_k\otimes
\Sigma_{k+1}e^{-\Sigma_k}\right) \nonumber\\[8pt]
&&
{1\over\chi_k}\left(\frac{\varsigma}{\chi_{k+1}}\right)^2\,\left(\Lambda^2_ke^{2\Sigma_{k}}\otimes\Sigma^2_{k+1}\Lambda_k
+\Lambda_k\otimes\Sigma^2_{k+1}e^{-\Sigma_k}
+\Sigma^2_{k+1}\otimes\Lambda_k
\right)
 \nonumber\\[8pt]&&
- {2\over\chi_k}\left(\frac{\varsigma}{\chi_{k+1}} \right)^2\,\Lambda_k \Sigma_{k+1} e^{\Sigma_k}\otimes\Sigma_{k+1}\Lambda_k\,
 \nonumber\end{eqnarray}
%$2H_k\Lambda_ke^{\Sigma_k}\otimes\Sigma_{k+1}\Lambda_k\,$
Here $\Sigma_{k+1}$ is denoted with index mod 2, i.e. $\Sigma_{k+1}$ is equal to
$\Sigma_0$ for $k=1$; further $\Lambda_k=e^{-2\Sigma_k}-e^{-\Sigma_k}=
-\chi_kE_{k+}e^{-2\Sigma_k}$
\footnote{One finds $[f(E_+),H]=-E_+f'(E_+)$, $[f(E_+),E_-]=2Hf'(E_+)-E_+f''(E_+)$, where $f$ is an analytic function of one variable. In particular
$[\Sigma,H]=-\chi E_+ e^{-\Sigma}=\Lambda e^{\Sigma}$, $[\Sigma,E_-]=2\chi He^{-\Sigma}-\chi\Lambda$.}.
Therefore, $\Lambda_k$ is proportional to $\chi_k$.

Using further the relations (\ref{blte7}-\ref{blte8}) one obtains the following formulae for the antipodes
 \begin{eqnarray}\label{blte22}
  S_{2}(E_{k+})&=&-E_{k+}\,e^{-\Sigma_k}  ,\qquad\qquad  S_{2}(H_k)=-H_ke^{\Sigma_k}\nonumber\\[8pt]
  S_{2}(E_{k-})&=&-E_{k-}\,e^{\Sigma_k} +\chi_k H^2_ke^{\Sigma_k}(e^{\Sigma_k}+1)
  -\chi_k^2 H_kE_{k+}e^{\Sigma_k}%(e^{\Sigma_k}-1)
  %\nonumber\\[8pt]
  %&-& (-)^k\,\frac{2\varsigma}{\chi_{k+1}}\,\Sigma_{k+1}\left(2e^{2\Sigma_k}-3e^{\Sigma_k}+e^{-\Sigma_k}\ \ -\chi^2_k\,\Sigma_{k+1}E_{k+}^2\right)\nonumber\\[8pt]
  %&-&2\chi_k\left(\frac{\varsigma}{\chi_{k+1}}\right)^2\,\Sigma^2_{k+1}E_{k+}^2 .
\end{eqnarray}
We notice that the Abelian twist (\ref{e12}) does not contribute to the antipodes (\ref{blte22}).
%The antipods for $\bar H,\, \bar E,\,\bar F$ are obtained by complex conjugation.

The quantum universal $R$-matrix $R_2\equiv R_{2^{}}^{}(\chi,\bar\chi,\varsigma)$ takes the form
\begin{equation}\label{qr2}
R_{2^{}}^{}=\exp{(\frac{-\varsigma}{\chi\bar\chi}\Sigma\wedge\bar\Sigma)}\exp{(\Sigma\otimes H)} \exp{(-H\otimes \Sigma)} \exp{(\bar\Sigma \otimes \bar H)}  \exp{(-\bar H\otimes \bar\Sigma)} \exp{(\frac{-\varsigma}{\chi\bar\chi}\Sigma\wedge\bar\Sigma)}.
\end{equation}
The formulae (\ref{e21})--(\ref{qr2}) present the general three-parameter deformation which can be studied in various 
%by executing a suitable limit to any combination of 
two-parameter limits. For example, if $\chi\mapsto 0$ one should  take
into account that $\lim_{\chi\mapsto 0}\frac{\Sigma}{\chi}=E_+$,  $\lim_{\chi\mapsto 0}\Lambda=0$ and
$\lim_{\chi\mapsto 0}\frac{\Lambda}{\chi}=-E_+$.
In this case the left chiral sector will be deformed only by Abelian twist.
The case $\varsigma=0$ provides obviously the product of two independent Jordanian deformations.

In real cases the independence of parameters may be not valid. Only for the real $\mathfrak{o}(2,2)$ deformation all three 
  parameters are imaginary and independent. In the Lorentzian case \footnote{Studied first time in \cite{BLT06}.} two Jordanian parameters $(\chi, \bar\chi)$ are replace by one as follows from the condition $\chi=(\bar\chi)^*$ in the table below. 
	%If $\chi=\bar\chi=0$ one obtains simple one parameter Abelian deformation of $\mathfrak{o}(3,1)$.
\begin{center}
% {\bf Table A. All real Lie bialgebra structures for $\mathfrak{o}(4;\mathbb{C})$ }\\[1ex]
\begin{table}[h]
\begin{tabular}{|c|c|c|c|}
\hline
&&&\\[-8pt]
$\mathfrak{o}''(2,2)$ &    $\chi,\bar\chi, \varsigma\in\imath\mathbb{R}$   &
$H^{\star}=-H, E_{\pm}^{\star}=-E_{\pm}$
& $\bar H^{\star}=-\bar H , \bar E_{\pm}^{\star}=-\bar E_{\pm}$\\
\hline
&&&\\[-8pt]
$\mathfrak{o}(3,1)$  &  $\chi=\bar\chi^*\in\imath\mathbb{R}$ ; $\varsigma\in\mathbb{R}$  & $H^{\ddagger} =-\bar H , E_{\pm}^{\ddagger}=-\bar E_{\pm}$
& $\bar H^{\ddagger}= - H  , \bar E_{\pm}^{\ddagger}=-E_{\pm}$\\
\hline
\end{tabular}
\caption{Real quantizations of
$r_{II}(\chi, \bar\chi, \varsigma)=\chi\,E_{+}\wedge H+\bar{\chi}\,\bar{E}_{+}\wedge\bar{H}+\varsigma E_{+}\wedge\bar{E}_{+}$}
	\label{tab:rr2}
\end{table}
\end{center}
All the twists present in the formula (\ref{e11}) are unitary (if the corresponding parameters are as indicated in the Table \ref{tab:rr2}) and the reality conditions (\ref{d-1}) are satisfied. 

\subsection{ Twisted pair of $q$-analogs ($r$-matrix $r_{III}$)}
  %$q$-deformation of Cartan type  for $\mathfrak{o}(4;\mathbb

%For the sake of convenience we introduce the following notations
%$z_{\pm}:=\beta\pm\imath\alpha$. It should be noted that $z_{-}^{}=z_{+}^*$ if the
%parameters $\alpha$ and $\beta$ are real, and  $z_{-}^{}=-z_{+}^*$ if the parameters
%$\alpha$ and $\beta$ are pure imaginary.

From the structure of the classical $r$-matrix $r_{3}$ (see (\ref{r3})) for $\eta=0$ follows that a quantum deformation 
$U_{r'_{3}} (\mathfrak{o}(3,1))$ is a combination of two independent $q$-analogs (stanadard deformations) of
$U(\mathfrak{sl}(2;\mathbb{C}))$ with the parameter $q=\exp{{1\over 2}\gamma}=q_{0}$ and $\bar q=\exp{{1\over 2} \bar{\gamma}}=q_{1}$.
Moreover one has the splitting $U_{(q,\bar q)}(\mathfrak{o}(4;\mathbb{C}))\cong U_{q_{}}(\mathfrak{sl}(2;\mathbb{C}))\otimes
U_{\bar q_{}}(\mathfrak{sl}(2;\mathbb{C}))$.% and the standard generators

This implies that the starting point for further considerations is a pair of standard (Drinfeld-Jimbo) deformations in  each chiral sector. They are described by nonlinear (quantum)  generators $q_{k^{}}^{\pm \H_{k}}$, $\E_{k\pm}\ (k=0,1)$ which %$q_{k^{}}^{\pm H_{2}}$, $E_{2\pm}$
satisfy the following defining relations
\begin{eqnarray}\label{ct1}
 q_{k}^{\H_k}\E_{k\pm}\!\!&=\!\!&q_{k}^{\pm 1}\E_{k\pm}\,q_{k}^{\H_k}~,\qquad
[\E_{k+},\,\E_{k-}]\;=\;\frac{q_{k}^{2\H_k}-q_{k}^{-2\H_k}}
{q_{k}^{}-q_{k}^{-1}}~,
%\\[7pt]\label{ct2}
%q_{z_{-}}^{H_2}E_{2\pm}\!\!&=\!\!&q_{z_{-}}^{\pm1}E_{2\pm}\,q_{z_{-}}^{H_2}~,\qquad
%[E_{2+},\,E_{2-}]\;=\;\frac{q_{z_{-}}^{2H_2}-q_{z_{-}}^{-2H_2}}
%{q_{z_{-}}^{}-q_{z_{-}}^{-1}}~.
\end{eqnarray}
The co-products $\Delta_{3'}$ and antipodes $S_{3'_{}}$ are given
by the formulas :
\begin{eqnarray}\label{ct3}
\Delta_{3'_{}}^{}(q_{k}^{\pm \H_{k}})\!\!&=\!\!&q_{k}^{\pm \H_{k}}\otimes
q_{k}^{\pm \H_{k}}~, \qquad\Delta_{3'_{}}^{}(\E_{k\pm})\;=\;\E_{k\pm}\otimes
q_{k}^{\H_{k}}+q_{k}^{-\H_{k}}\otimes \E_{k\pm}~,
%\\[7pt]\label{ct4}
%\Delta_{r'_{3}}^{}(q_{z_{-}}^{\pm H_{2}})\!\!&=\!\!&q_{z_{-}}^{\pm H_{2}}\otimes
%q_{z_{-}}^{\pm H_{2}}~,\qquad\Delta_{r'_{3}}^{}(E_{2\pm})\;=\;E_{2\pm}\otimes
%q_{z_{-}}^{H_{2}}+q_{z_{-}}^{-H_{2}}\otimes E_{2\pm}~,
\end{eqnarray}
\begin{eqnarray}\label{ct5}
S_{3'_{}}^{}(q_{k}^{\pm \H_{k}})\!\!&=\!\!&q_{k}^{\mp \H_{k}}~,\qquad
S_{3'_{}}^{}(\E_{k\pm})\;=\;-q_{k}^{\pm1}\E_{k\pm}~,
%\\[7pt]\label{ct6}
%S_{r'_{3}}^{}(q_{z_{-}^{}}^{\pm H_{2}})\!\!&=\!\!&q_{z_{-}}^{\mp H_{2}}~,\qquad
%S_{r'_{3}}^{}(E_{2\pm})\;=\;-q_{z_{-}}^{\,\mp1}E_{2\pm}~.
\end{eqnarray}

The universal $R$-matrices $R_{3'k}$ for each chiral sector
are well-known and using deformed CW generators (\ref{ct1}) take the form ($q_k=\exp{{1\over 2}\gamma_k}$):
\begin{eqnarray}\label{ct13}
R_{3'k}(\gamma_k)\!\!&=\!\!&\exp_{q_{k}^{-2}}\Bigl((q_{k}^{}-q_{k}^{-1})\E_{k+}\,
q_{k}^{-\H_{k}}\otimes q_{k}^{\H_{k}}\E_{k-}\Bigr) q_{k}^{2\H_{k}\otimes \H_{k}}~,
%% \\[15pt]
\end{eqnarray}
Following the discussion of nontriangular case in Sect. 2.1, there exists alternative universal $R$-matrix in the form
\begin{eqnarray}\label{ct13bis}
(R_{3'k}^{\tau})^{ -1}\!\!&=\!\!&q_{k}^{-2\H_{k}\otimes \H_{k}}\,\exp_{q_{k}^{2}}\Bigl((q_{k}^{-1}-q_{k}^{})q_{k}^{\H_{k}}\E_{k-}\otimes \E_{k+}\,q_{k}^{-\H_{k}}\Bigr) ~.
%% \\[15pt]
\end{eqnarray}
%as well instead of (\ref{ct13}).

Therefore, the universal $R$-matrix $R_{3'_{}}$, which connects the coproducts
$\Delta_{3'_{}}^{12}:=\Delta_{3'_{}}^{}$ and the flipped one $\Delta_{3'_{}}^{21}$  
can be  written in two equivalent forms: \footnote{In fact, taking into account (\ref{ct13bis}), there are four 
ways of describing  universal $R$- matrix $R_{3'}$.}
\begin{eqnarray}\label{ct12}
R_{3'_{}}(\gamma,\bar\gamma)\!\!&=\!\!&R_{3'0}(\gamma)R_{3'1}(\bar\gamma)\,=\,R_{3'1}(\bar\gamma)R_{3'0}(\gamma)~,
\end{eqnarray}
Expanding (\ref{ct12}) up to first order in deformation parameters $(\gamma,\bar\gamma)$ one gets
\begin{eqnarray}\label{ct18}
R_{3'_{}}(\gamma,\bar\gamma)\!\!&=\!\!&1+r_{3'BD}^{}+O(\gamma^2,\gamma\bar\gamma,\bar\gamma^2)~,
\end{eqnarray}
where $r_{3'BD}^{}$ is in Belavin-Drinfeld form \footnote{In (\ref{ct19}) and in other formulas describing  classical $r$-matrices, %(\ref{ct22})
the generators $E_{\pm}$, $\bar E_{\pm}$ are not deformed.}
\begin{eqnarray}\label{ct19}
r_{3'BD}^{}\!\!&=\!\!&\gamma\bigl(E_{+}\otimes E_{-}+H\otimes H\bigr)
+\bar\gamma\bigl(\bar E_{+}\otimes \bar E_{-}+\bar H\otimes \bar H\bigr)
\end{eqnarray}
This $r$-matrix is not skew-symmetric
%\footnote{Notice that skew-symmetric part gives exactly the classical $r-$matrix ().},
and satisfies  the condition
\begin{eqnarray}\label{ct21}
r_{BD}^{12}+r_{BD}^{21}\!\!&=\!\!&\omega
\end{eqnarray}
where $\omega$ is the quadratic split  Casimir %or Cartan-Killing non-degenerate form
of $\mathfrak{o}(4; \mathbb{C})$
\begin{equation}
\begin{array}{rcl}\label{ct22}
\omega\!\!&=\!\!&\gamma \bigl(E_{+}\otimes E_{-}+E_{-}\otimes E_{+}+2H\otimes
H\bigr)
\\[7pt]
&&\!\!+\bar\gamma \bigl(\bar E_{+}\otimes \bar E_{-}+\bar E_{-}\otimes \bar E_{+}+2\bar H\otimes \bar H\bigr)
 \end{array}
\end{equation}
We recall that the Belavin-Drinfeld $r$-matrix $r_{BD}^{}$ satisfies CYBE and the $r$-matrix $r_3'$ is a skew-symmetric 
part of it. %$r_{BD}^{}$,
%namely.
%the homogeneous classical
%Yang-Baxter equation and the $r$-matrix $r_3'$ is a skew-symmetric part of $r_{BD}^{}$,
%namely
%\begin{eqnarray}\label{ct23}
%r_{BD}^{}\!\!&=\!\!& \frac{1}{2}\,r'_3+\frac{1}{2}\,\omega~.
%\end{eqnarray}
%Of course, switching off one of the deformation parameters $(\gamma, \bar\gamma)$ we leave one of the chiral sectors  not deformed.

Now we consider deformation of the quantum algebra   $U_{(\gamma, \bar\gamma)}(\mathfrak{o}(4;\mathbb{C}))\cong
U_{\gamma}(\mathfrak{sl}(2;\mathbb{C}))\otimes U_{\bar\gamma}(\mathfrak{sl}(2;\mathbb{C}))$ generated by the
$r$-matrix $r''_{3}=\eta H\otimes\bar H$, (see (\ref{r3})). Since the generators $\H$ and $\bar \H$ have the
primitive coproduct
\begin{eqnarray}\label{ct24}
\Delta_{3'_{}}(\H_{k})\!\!&=\!\!&\H_{k}\otimes 1+1\otimes \H_{k}\quad(k=0, 1)~,
\end{eqnarray}
the  Abelian two-tensor ($\tilde q=\exp{{1\over 4}\eta}$)
\begin{eqnarray}\label{ct25}
F_{3''}(\eta)\!\!:=\!\!&\tilde q_{}^{\H\wedge\bar \H}\qquad
%(F_{r_{3}''}^*\;=\;F_{r_{3}''}^{-1})
\end{eqnarray}
satisfies the 2-cocycle condition (\ref{coc}). Thus the complete deformation generated by the
$r$-matrix $r_{3}^{}$ is the twist deformation of $U_{(\gamma,\bar\gamma)}(\mathfrak{o} (4;\mathbb{C}))$;
the resulting coproduct $\Delta_{3_{}}^{}$ is given as follows
\begin{eqnarray}\label{ct28}
\Delta_{3_{}^{}}^{}(a)\!\!&=\!\!&F_{3''}^{}\Delta_{3'}^{}(a)
F_{3''}^{-1}\quad(\forall a\in U_{r'_{3}}(\mathfrak{o}(4;\mathbb{C}))~,
\end{eqnarray}
and the antipode $S_{3^{}}^{}$ is not changed ($S_{3^{}}^{}=S_{3'_{}}^{}$). Applying the twist (\ref{ct25}) to the
formulas (\ref{ct3})  we obtain
\begin{eqnarray}\label{ct29}
\Delta_{3_{}^{}}(q_{k}^{\pm \H_{k}})\!\!&=\!\!&q_{k}^{\pm \H_{k}}\otimes
q_{k}^{\pm \H_{k}},%\quad\Delta_{r_{3}}(q_{z_{-}}^{\pm H_{2}})\;=\;q_{z_{-}}^{\pm
%H_{2}}\otimes q_{z_{-}}^{\pm H_{2}},
\\[7pt]\label{ct30}
\Delta_{3_{}^{}}(\E_{k\pm})\!\!&=\!\!&\E_{k\pm}\otimes q_{k}^{\H_{k}}
\tilde q_{}^{\pm (-)^k\H_{k+1}}+q_{k}^{-\H_{k}}\tilde q_{}^{\mp (-)^k \H_{k+1}}\otimes \E_{k\pm}~.
%\\[7pt]\label{ct31}
%\Delta_{r_{3}}(E_{2\pm})\!\!&=\!\!&E_{2\pm}\otimes q_{z_{-}}^{H_{2}}
%q_{\imath\gamma}^{\mp H_{1}}+q_{z_{-}}^{-H_{2}}q_{\imath\gamma}^{\pm H_{1}}\otimes
%E_{2\pm}~.
\end{eqnarray}
The universal $R$-matrix, $R_{3_{}^{}}(\gamma,\bar\gamma,\eta)$, corresponding to the complete $r$-matrix
$r_{3}^{}$, has the form
\begin{eqnarray}%\label{ct32}
R_{3_{}^{}}(\gamma,\bar\gamma,\eta)\!\!&=\!\!&\tilde q_{}^{\bar \H_{}\wedge \H_{}}R_{3'_{}}(\gamma,\bar\gamma)
\tilde q_{}^{\bar \H_{}\wedge \H_{}}\;=\;R_{30}^{}(\gamma,\eta)R_{31}^{}(\bar\gamma,\eta) \tilde q_{}^{2\bar \H_{}\wedge
\H_{}}\;= \;R_{31}^{}(\bar\gamma,\eta)R_{30}^{}(\gamma,\eta) \tilde q_{}^{2\bar \H_{}\wedge \H_{}}~,\nonumber
\label{ct32}
\end{eqnarray}
where
\begin{eqnarray}\label{ct33}
R_{3k}^{}(\gamma_k,\eta)\!\!&=\!\!&\exp_{q_{k}^{-2}}\Bigl((q_{k}^{}-q_{k}^{-1})\E_{k+}
q_{k}^{-\H_{k}} \tilde q_{}^{(-)^{k+1}\H_{k+1}}\otimes q_{k}^{\H_{k}}
\tilde q_{}^{(-)^{k+1}\H_{k+1}} \E_{k-}\Bigr)\,q_{k}^{2\H_{k}\otimes \H_{k}}
\end{eqnarray}
%and
%\begin{eqnarray}\label{ct34}
%R_{2}\!\!&=\!\!&\exp_{q_{z_{-}}^{-2}}\Bigl((q_{z_{-}}-q_{z_{-}}^{-1})
%E_{2+}q_{z_{-}}^{-H_{2}}q_{\imath\gamma}^{H_{1}}\otimes q_{z_{-}}^{-H_{2}}
%q_{\imath\gamma}^{H_{1}}E_{2-}\Bigr)\,q_{z_{-}}^{2H_{2}\otimes H_{2}}
%\end{eqnarray}
In the linear limit   we obtain (cf. (\ref{ct12}),
(\ref{ct18}))
\begin{eqnarray}\label{ct37}
R_{3_{}^{}}\!\!&\sim\!\!&1+ r_3 %{1\over 2} r_3^{}+\,{1\over 2} \omega~.
\end{eqnarray}
This deformation admits seven real forms  which employ all four conjugations (cf. (\ref{rf1}) -- (\ref{rf4}) )~. 
The list of real forms  with corresponding restricted values of the deformation parameters $\gamma,\bar\gamma,\eta$ is 
presented in the Table \ref{tab:rr3}, with real bialgebras denoted in the first column (cf. Table \ref{tab:Real}).
%Below we remind how the corresponding  conjugations act on the Hopf algebra generators as anti-linear involutive antiautomorphisms. It should be noted that in all realizations, except the Lorentz one, elements $q^\H, q^{\bar \H}$ remain selconjugated.
\begin{center}
% {\bf Table A. All real Lie bialgebra structures for $\mathfrak{o}(4;\mathbb{C})$ }\\[1ex]
\begin{table}[h]
\begin{tabular}{|c|c|c|c|c|}
\hline
&&&\\[-8pt]
$\mathfrak{o}(4)$ & $\gamma,\bar\gamma\in\mathbb{R}$ ; $\eta\in\imath\mathbb{R}$& $(q^\H)^{\dag}=q^\H, \E_{\pm}^{\dag}=\E_{\mp}$
& $(\bar q^{\bar\H})^{\dag}=\bar q^{\bar{\H} }, \bar\E_{\pm}^{\dag}=\bar\E_{\mp}$& R\\
\hline
&&&\\[-8pt]
$\mathfrak{o}^\star(4)$ & $\gamma,\bar\gamma\in\mathbb{R}$ ; $\eta\in\imath\mathbb{R}$& $(q^\H)^{\dag}=q^\H, \E_{\pm}^{\dag}=\E_{\mp}$
& $(\bar q^{\bar\H})^{\#}=q^{\bar\H} , \bar\E_{\pm}^{\#}=-\bar\E_{\mp}$& R\\
\hline
&&&\\[-8pt]
$\mathfrak{o'}^{\star}(4)$ &  $\gamma,\eta\in\mathbb{R}$ ; $\bar\gamma\in\imath\mathbb{R}$  &$(q^\H)^{\dag}=q^\H, \E_{\pm}^{\dag}=\E_{\mp}$
& $(\bar q^{\bar\H})^{\star}=\bar q^{\bar\H} , \bar\E_{\pm}^{\star}=-\bar\E_{\pm}$& H\\
\hline
&&&\\[-8pt]
$\mathfrak{o}(2,2)$ & $\gamma,\bar\gamma\in\mathbb{R}$ ; $\eta\in\imath\mathbb{R}$  &  $(q^\H)^{\#}=q^\H, \E_{\pm}^{\#}=-\E_{\mp}$
& $(\bar q^{\bar\H})^{\#}=\bar q^{\bar\H} , \bar\E_{\pm}^{\#}=-\bar\E_{\mp}$& R\\
\hline
&&&\\[-8pt]
$\mathfrak{o}'(2,2)$ &  $\gamma,\eta\in\mathbb{R}$ ; $\bar\gamma\in\imath\mathbb{R}$ &$(q^\H)^{\#}=q^\H, \E_{\pm}^{\#}=-\E_{\mp}$
& $(\bar q^{\bar\H})^{\star}=\bar q^{\bar\H} , \bar\E_{\pm}^{\star}=-\bar\E_{\pm}$& H\\
\hline
&&&\\[-8pt]
$\mathfrak{o}''(2,2)$ &    $\gamma,\bar\gamma, \eta\in\imath\mathbb{R}$   &
$(q^\H)^{\star}=q^\H, \E_{\pm}^{\star}=-\E_{\pm}$
& $(\bar q^{\bar\H})^{\star}=\bar q^{\bar\H} , \bar\E_{\pm}^{\star}=-\bar\E_{\pm}$& A\\
\hline
&&&\\[-8pt]
$\mathfrak{o}(3,1)$  &  $\bar\gamma=-\gamma^*\in\mathbb{C}$ ; $\eta\in\mathbb{R}$  & $(q^\H)^{\ddagger} =\bar{q}^{\bar\H} , \E_{\pm}^{\ddagger}=-\bar\E_{\pm}$
& $(\bar q^{\bar\H})^{\ddagger}=q^{\H}  , \bar\E_{\pm}^{\ddagger}=-\E_{\pm}$& A\\
\hline
\end{tabular}
\caption{Real quantizations of $r_{III}(\gamma, \bar\gamma, \eta)=
\gamma\,E_{+}\wedge E_{-} +\bar{\gamma}\,\bar{E}_{+}\wedge\bar{E}_{-} + \eta\,H\wedge\bar{H}
$}
	\label{tab:rr3}
\end{table}
\end{center}
The letters in the last column (R=real, A=antireal, H=hybrid) indicate the properties of the $R$-matrix under respective conjugation:
$R^\star=R^\tau$ for real, $R^\star=R^{-1}$ for antireal cases.
In the hybrid case the $R$-matrix decomposes into a product of three factors, with first real, second antireal and  third is given by twist which satisfies both reality conditions.

It shoul be mentioned that only the classical $r$-matrix $r_{III}$ provides the quantum deformations of real $\mathfrak{o}(4)= \mathfrak{su}(2)\oplus \mathfrak{su)(2}$ algebra (see first line in Tab \ref{tab:rr3}). Particular case, with $\eta=0$ was derived as describing quantum symmetries of $D=3$ LQG \cite{JKG}.

\subsection{ Twisting of  $\mathfrak{o}(4;\mathbb{C})$
%of $\mathfrak{o}(4;\mathbb{C})$, corresponding
Belavin--Drinfeld triple
 ($r$-matrix $r_{IV}$)}

Next, we describe quantum deformation corresponding to the classical $r$-matrix $r_{IV}$
(\ref{r4}). Since the $r$-matrix $r_{IV}(\gamma,0):=r_{4'}$ is a particular case of
$r_{III}^{}(\gamma,\bar\gamma,\eta)$, namely
$r_{IV'}(\gamma)=r_{III}^{}(\gamma,-\gamma,-2\gamma)$, $\gamma\in\mathbb{C}$, the quantum deformation
corresponding to the $r$-matrix $r_{IV'}$ is obtained from the formulae in Sect. 5.3  by setting
$\bar q=\tilde q= q^{-1}$. The quantum deformation corresponding to $r_{IV}$ %$U_{r_{4'}}(\mathfrak{o}(3;\mathbb{C}))$ 
is generated by the elements $q_{}^{\pm \H_{k}}$, $\E_{k\pm}$ (k=0,1) %and $q_{}^{\pm \bar \H_{}}$,$\bar \E_{\pm}$ 
with the following defining relations (cf. (\ref{ct1}))
\begin{eqnarray}\label{BD1}
q_{}^{\H_k}\E_{k\pm}\!\!&=\!\!&q_{}^{\pm1}\E_{k\pm}\,q_{}^{\H_k}~,\qquad
[\E_{k+},\,\E_{k-}]\;=\;\frac{q_{}^{2\H_k}-q_{}^{-2\H_k}}
{q_{}^{}-q_{}^{-1}}%\quad\;\; (k=0,1)~.
\end{eqnarray}
constituting the algebra $U_{q_{}}(\mathfrak{sl}(2;\mathbb{C}))\otimes U_{ q^{-1}}(\mathfrak{sl}(2;\mathbb{C}))$. 
The co-products $\Delta_{4'_{}}$ and antipodes $S_{4'_{}}$ generated by $r_{IV'}$ are given by the formulas 
(cf. (\ref{ct29})-(\ref{ct30})):
\begin{eqnarray}\label{BD2}
\Delta_{4'_{}}(q_{}^{\pm \H_{k}})\!\!&=\!\!&q_{}^{\pm \H_{k}}\otimes q_{}^{\pm
\H_{k}} ~,\nonumber
\\[7pt]\label{BD3}
\Delta_{4'_{}}(\E_{k\pm})\!\!&=\!\!&\E_{k\pm}\otimes q_{}^{(-)^k(\H_{k}\pm \H_{k+1})}+
q_{}^{(-)^{k+1}(\H_{k}\pm \H_{k+1})}\otimes \E_{k\pm}~,
\\[7pt]\label{BD4}
%\Delta_{r_{4}'}(E_{2\pm})\!\!&=\!\!&E_{2\pm}\otimes q_{\xi}^{\mp H_{1}-H_{2}}+
%q_{\xi}^{\pm H_{1}+H_{2}}\otimes E_{2\pm}~,
%\\[7pt]\label{BD5}
S_{4'_{}}^{}(q_{}^{\pm \H_{k}})\!\!&=\!\!&q_{}^{\mp \H_{k}} ,\qquad
%\\[7pt]\label{BD6}
S_{4'_{}}^{}(\E_{k\pm})\!\!=\!\!-q_{}^{\pm(-)^k}\E_{k\pm}~,\nonumber
%S_{r'_{4}}^{}(E_{2\pm})\;=\;-q_{\xi}^{\,\mp1}E_{2\pm}~.
\end{eqnarray}
The full  deformation of the quantum algebra (\ref{BD1})--(\ref{BD2})
%$U_{q}(\mathfrak{o}(4;\mathbb{C}))$ (secondary quantization %of $U(\mathfrak{o}(3,1))$)
is obtained after performing the twist quantization generated by
the remaining part of the  $r$-matrix $r_{IV}$ namely $r_{IV''}=\varsigma E_+\wedge\bar E_+$,  described
by the following quantum Abelian twist factor \cite{IO01}:
\begin{eqnarray}\label{BD7}
F_{4_{}''}(\gamma,\varsigma)\!\!&:=\!\!&\exp_{q_{}^{2}}^{}\big(\varsigma \E_{+}q_{}^{\H_{}+\bar \H_{}} \otimes
q_{}^{\H_{}+\bar \H_{}}\bar \E_{+}\big)~.
\end{eqnarray}
It can be shown that the two-tensor (\ref{BD7})
%is unitary, $F_{r_{4}''}^*=F_{r_{4}''}^{-1}$, with
%respect to the $*$-flipped involution. Moreover, using properties of $q$-exponentials
%(see \cite{KhTo1}) is not hard to verify that $F_{r_{4}''}$
satisfies the 2-cocycle equation (\ref{coc}). 
%Thus the quantization corresponding to the $r$-matrix $r_4$ is the twisted $q$-deformation $U_{r_{4}'}(\mathfrak{o}(4;\mathbb{C}))$.

Explicit form of the co-products $\Delta_{4_{}^{}}^{}(\cdot)= F_{4_{}''}^{}
\Delta_{4_{}'}^{}(\cdot)F_{4_{}''}^{-1}$ in the complex Cartan-Weyl bases of
$U_{r_{4}'}(\mathfrak{o}(4;\mathbb{C}))$ can be calculated using $q$-analog of Hadamard formula (Appendix B)
\begin{eqnarray}\label{BD8}
\Delta_{4_{}}(q_{}^{\pm(\H_{}-\bar \H)})\!\!&=\!\!&q_{}^{\pm(\H_{}-\bar \H)}\otimes
q_{}^{\pm(\H_{}-\bar \H)}~,\nonumber
\\[5pt]\label{BD9}
\Delta_{4_{}}(\,q_{}^{\;\,\H_{}+\bar \H})\!\!&=\!\!&\mathbb{X}^{-1}\,
q_{}^{\H_{}+\bar \H}\otimes q_{}^{\H_{}+\bar \H}~,\nonumber
\\[7pt]\label{BD10}
\Delta_{4_{}}(q_{}^{\;-\H_{}-\bar \H})\!\!&=\!\!&q_{}^{\,-\H_{}-\bar \H}\otimes
q_{}^{-\H_{}-\bar \H}\,\mathbb{X}~,\nonumber
\\[7pt]\label{BD11}
\Delta_{4_{}}(\E_{+})\!\!&=\!\!&\E_{+}\otimes q_{}^{\H_{}+\bar \H_{}}+
q_{}^{-\H_{}-\bar \H_{}} \otimes \E_{+}\,\mathbb{X}~,\nonumber
\\[7pt]\label{BD12}
\Delta_{4_{}}(\bar \E_{+})\!\!&=\!\!&\bar \E_{+}\otimes q_{}^{-\H_{}-\bar \H_{}}\mathbb{X}+
q_{}^{\H_{}+\bar \H_{}}\otimes \bar \E_{+}~,
\end{eqnarray}
\begin{eqnarray}\label{BD13}
\begin{array}{rcl}
\Delta_{4_{}}(\E_{-})\!\!&=\!\!&\E_{-}\otimes q_{}^{\H_{}-\bar \H_{}}+
q_{}^{\bar \H_{}-\H_{}}\otimes \E_{-}\,-
\\[12pt]
&&-\;\displaystyle\frac{\varsigma}{q_{}^{}-q_{}^{-1}}\,\bigl(q_{}^{-4\H_{}}\otimes1-
\mathbb{X}^{-1}\bigr)\bigl(q_{}^{3\H_{}+\bar \H_{}}\otimes \bar \E_{+}q_{}^{2\H}\bigr)~,
\end{array}\nonumber
\end{eqnarray}
\begin{eqnarray}\label{BD14}
\begin{array}{rcl}
\Delta_{4_{}}(\bar \E_{-})\!\!&=\!\!&\bar \E_{-}\otimes q_{}^{\H_{}-\bar \H_{}}+
q_{}^{\bar \H_{}-\H_{}}\otimes \bar \E_{-}\,-
\\[12pt]
&&-\;\displaystyle\frac{\varsigma}{q_{}^{}-q_{}^{-1}}\,\bigl(1\otimes q_{}^{-4\bar \H_{}}-
\mathbb{X}^{-1}\bigr)\bigr(\E_{+}q_{}^{2\bar \H}\otimes q_{}^{\H_{}+3\bar \H_{}}\bigr)~,
\end{array}\nonumber
\end{eqnarray}
where
\begin{eqnarray}\label{BD15}
\mathbb{X}\!\!&:=\!\!&1+\varsigma (q_{}^{2}-1)\E_{+}q_{}^{\H_{}+\bar \H_{}}\otimes
q_{}^{\H_{}+\bar \H_{}} \bar \E_{+}~.
\end{eqnarray}
Explicit formulas for antipodes $S_{4}(\cdot)=uS_{4_{}'}(\cdot)u^{-1}$ where
\begin{eqnarray}\label{BD16}
u^{-1}\!\!&=\!\!&m\circ (S_{4_{}'}\otimes\mathop{\rm
id})\exp_{q_{}^{2}}^{}\big(\varsigma\,\E_{+}q_{}^{\H_{}+\bar \H_{}}\otimes
q_{}^{\H_{}+\bar \H_{}}\bar \E_{+}\big)\,=\,\exp_{q_{}^{2}}^{}\big(\!\varsigma\,
\E_{+}\bar \E_{+}\big)~,
\end{eqnarray}
are given  (as results from $q$-Hadamard formula) below
\begin{eqnarray}\label{BD17}
S_{4_{}}(q_{}^{\pm(\H_{}-\bar \H)})\!\!&=\!\!&q_{}^{\mp(\H_{}-\bar\H)}~,\qquad
S_{4_{}}(\E_{k+})\!\!=\!\!-q_{}^{(-)^k}\E_{k+} \,,\nonumber
%S_{r_{4}}(q_{\xi}^{-(H_{1}-H_2)})\;=\;q_{\xi}^{H_{1}-H_2}~,
\\[7pt]\label{BD18}
S_{4_{}}(\,q_{}^{\H_{}+\bar \H})\!\!&=\!\!&q_{}^{-\H_{}-\bar \H}X^{-1}~,\quad
S_{4_{}}(q_{}^{\;-\H_{}-\bar \H})\;=\;X\,q_{}^{\H_{}+\bar \H}~,
\\[7pt]\label{BD19}
%S_{4_{}}(E_{k+})\!\!&=\!\!&-q_{}^{(-)^k}E_{k+}~,\qquad\qquad\quad\;
%S_{r_{4}}(E_{2+})\;=\;-q_{\xi}^{-1}E_{2+}~,
%\\[7pt]\label{BD20}
S_{4_{}}(\E_{k-})\!\!&=\!\!&-q_{}^{(-1)^{k+1}}\E_{k-}+\frac{(-)^k\varsigma}{q_{}^{2(-)^k}-1}\,
\E_{(k+1)+}\bigl(q_{}^{2\H_k}-q_{}^{-2\H_k}X^{-1}\bigr)~,\nonumber
%\\[7pt]\label{BD21}
%S_{r_{4}}(E_{2-})\!\!&=\!\!&-q_{\xi}^{}E_{2-}-\frac{\eta}{q_{\xi}^{-2}-1}\,E_{1+}q_{\xi}^{-2H_2}
%\bigl(q_{\xi}^{4H_2}-X^{-1}\bigr)~,
\end{eqnarray}
where
\begin{eqnarray}\label{BD22}
X\!\!&:=\!\!&1+\varsigma(q_{}^{2}-1)\E_{+}\bar \E_{+}~.
\end{eqnarray}
Therefore a total universal $R$-matrix for this case is the following product 
(now $\bar\gamma=-\gamma \Leftrightarrow\bar q=q^{-1}$)
\begin{eqnarray}\label{ct12b}
R_{4_{}}(\gamma,\varsigma)\!\!&=\!\!&F^\tau_{4_{}''}(\gamma,\varsigma)R_{3'0}(\gamma)R_{3'1}(-\gamma)
F^{-1}_{4_{}''}(\gamma,\varsigma)~.
%\,=\,R_{4'1}R_{4'0}
\end{eqnarray}
Two real qunatizations are described in the Table \ref{tab:rr4}.

\begin{center}
% {\bf Table A. All real Lie bialgebra structures for $\mathfrak{o}(4;\mathbb{C})$ }\\[1ex]
\begin{table}[h]
\begin{tabular}{|c|c|c|c|c|}
\hline
&&&\\[-8pt]
$\mathfrak{o}''(2,2)$ & \ \ \ \; $\gamma,\,\varsigma\in\imath\mathbb{R}\ \ \ $  &  $(q^\H)^{\star}=q^\H, \E_{\pm}^{\star}=-\E_{\pm}$
& $(\bar q^{\bar\H})^{\star}=\bar q^{\bar\H} , \bar\E_{\pm}^{\star}=-\bar\E_{\pm}$& A\\
\hline
&&&\\[-8pt]
$\mathfrak{o}(3,1)$  &  $\ \ \ \ \gamma\in\mathbb{R}\ \ ,\ \varsigma=0 $  & $(q^\H)^{\ddagger} =q^{-\bar\H} , \E_{\pm}^{\ddagger}=-\bar\E_{\pm}$
& $(q^{\bar\H})^{\ddagger}=q^{-\H}  , \bar\E_{\pm}^{\ddagger}=-\E_{\pm}$& A\\
\hline
\end{tabular}
\caption{Real quantizations of $r_{IV}(\gamma,\varsigma)=
\gamma\left(E_{+}\wedge E_{-} -\bar{E}_{+}\wedge\bar{E}_{-} -2H\wedge\bar{H}\right)+\varsigma E_{+}\wedge\bar{E}_{+}
$}
	\label{tab:rr4}
\end{table}
\end{center}
It should be noted that the value $\varsigma=0$ in the Lorentzian case is due to the fact that twist (\ref{BD7}), in contrast to the $\mathfrak{o}''(2,2)$ case where $|q|=1$,  is not unitary for real $q$. In order to have formulae (\ref{BD8})--
(\ref{BD17}) compatible with the Lorentzian conjugation (\ref{rf4}) it is helpful to introduce flipped conjugation (\ref{d-1f}) on the tensor product of quantized algebras (see \cite{BLT08}).

Alternatively, one can keep the standard  (non-flipped) conjugation and seek for the unitarizing coboundary twist - the quantum analog of 
(\ref{Atwist2}) \footnote{This method has been e.g. used in \cite{BLT03} in order to unitarize superextension of the Jordanian deformation.}. Examples of quantum coboundary twists can be found e.g. in  \cite{Sam06}. The realization of this task is postponed to our future work.

Yet another method relying on quantum deformation of the real involution  ($\star$-involution) has been studded in 
\cite{Lyub90,Maj95,Osei}. %We shall apply it now. 
Assuming $q$ real, the quantum twist (\ref{BD7}) in the real Hopf algebra 
$(U_{q_{}}(\mathfrak{sl}(2;\mathbb{C}))\otimes U_{ q^{-1}}(\mathfrak{sl}(2;\mathbb{C}), \Delta_{4'}, S_{4'}, \ddag)$ satisfies the condition (see \cite{Maj95}, Prop. 2.3.7, p.59)
 \begin{equation}
(S_{4'}^{}\otimes S_{4'})(F_{4''}^{\ddag\otimes\ddag})=F_{4''}^\tau
\nonumber\end{equation}
for $\varsigma$ real. This permits to introduce new conjugation quantum-deformed by the similarity transformation  
\begin{equation}
()^{\ddag^\prime} = u\,()^\ddag\,u^{-1}
\nonumber\end{equation}
where $u^{-1}$ is given by the formula (\ref{BD16})
(in our case $S^{-1}(u)=u^{}$). 
Explicit calculations with the help of $q$-Hadamard formula leads to the following results 
\begin{eqnarray}\label{BDstar}
(q_{}^{\pm(\H_{}-\bar \H)})^{\ddag'}\!\!&=\!\!&(q_{}^{\pm(\H_{}-\bar \H)})^{\ddag}=q_{}^{\pm(\H_{}-\bar\H)}~,\qquad
(\E_{k+})^{\ddag'}=(\E_{k+})^{\ddag}=-\E_{(k+1)+} \,,\nonumber
%S_{r_{4}}(q_{\xi}^{-(H_{1}-H_2)})\;=\;q_{\xi}^{H_{1}-H_2}~,
\\[7pt]
(\,q_{}^{\H_{}+\bar \H})^{\ddag'}\!\!&=\!\!&q_{}^{-\H_{}-\bar \H}X^{-1}~,\quad
(q_{}^{\;-\H_{}-\bar \H})^{\ddag'}\;=\;X^{}\,q_{}^{\H_{}+\bar \H}~,
\\[7pt]
%S_{4_{}}(E_{k+})\!\!&=\!\!&-q_{}^{(-)^k}E_{k+}~,\qquad\qquad\quad\;
%S_{r_{4}}(E_{2+})\;=\;-q_{\xi}^{-1}E_{2+}~,
%\\[7pt]\label{BD20}
(\E_{k-})^{\ddag'}\!\!&=\!\!&-\E_{(k+1)-}+\frac{\varsigma}{q^{}-q^{-1}}\,
\E_{k+}\bigl(q_{}^{2\H_{k+1}}-q_{}^{-2\H_{k+1}}X^{-1}\bigr)~,\nonumber
%\\[7pt]\label{BD21}
%S_{r_{4}}(E_{2-})\!\!&=\!\!&-q_{\xi}^{}E_{2-}-\frac{\eta}{q_{\xi}^{-2}-1}\,E_{1+}q_{\xi}^{-2H_2}
%\bigl(q_{\xi}^{4H_2}-X^{-1}\bigr)~,
\end{eqnarray}
where $X$ is given by (\ref{BD22}). 
In this way Belavin-Drinfeld type quantum deformation of the Lorentz algebra is described by the real Hopf algebra
$(U_{q_{}}(\mathfrak{sl}(2;\mathbb{C}))\otimes U_{ q^{-1}}(\mathfrak{sl}(2;\mathbb{C})[[\varsigma]], \Delta_{4}, S_{4}, \ddag')$.

\subsection{ Left $q$-analog and right Jordanian deformation intertwined by Abelian twist ($r$-matrix $r_{V}$)}
%Mixed  standard -- Jordanian -- Abelian deformations  of $\mathfrak{o}(4;\mathbb{C})$  }

In this case we start with the left sector as q-deformed with $q=\exp{{1\over 2}\gamma}$ 
\begin{eqnarray}\label{BD}
q_{}^{\H}\E_{\pm}\!\!&=\!\!&q_{}^{\pm1}\E_{\pm}\,q_{}^{\H}~,\qquad
[\E_{+},\,\E_{-}]\;=\;\frac{q_{}^{2\H}-q_{}^{-2\H}}
{q_{}^{}-q_{}^{-1}}\quad\;\; ~,
\end{eqnarray}
the right sector is deformed by Jordanian twist $F_J$ expressed in undeformed CW basis (cf. Sect. 3.1.1)
\begin{eqnarray}\label{pr16}
[\bar{H},\,\bar{E}_{\pm}]\!\!&=\!\!&\bar{E}_{\pm}~,\quad [\bar{E}_{+},\,\bar{E}_{-}]\,=\,2\bar{H}~.
\end{eqnarray}
%Then, having determined the algebraic sector (\ref{BD})-(\ref{pr16}),  
Further we perform the subsequent quantization by using the quantized Abelian twist
$$F_{5''}(\bar\chi,\rho)=\tilde q^{\H\wedge\bar\Sigma} ,\qquad \tilde q=\exp{{\rho\over 4\bar\chi}}$$
The explicit coproduct formuale   are the following
\begin{eqnarray}\label{ct29b}
\Delta_{5_{}^{}}(q_{}^{\pm \H_{}})\!\!&=\!\!&q_{}^{\pm \H_{}}\otimes
q_{}^{\pm \H_{}},%\quad\Delta_{r_{3}}(q_{z_{-}}^{\pm H_{2}})\;=\;q_{z_{-}}^{\pm
%H_{2}}\otimes q_{z_{-}}^{\pm H_{2}},
\\[7pt]\label{ct30b}
\Delta_{5_{}^{}}(\E_{\pm})\!\!&=\!\!&\E_{\pm}\otimes q_{}^{\H_{}}
\tilde q_{}^{\pm \bar\Sigma}+q_{}^{-\H_{}}\tilde q_{}^{\mp \bar\Sigma }\otimes \E_{\pm}~.
\end{eqnarray}
\begin{eqnarray}\label{blte21}
\Delta_{5}(\bar E_{+})
&=&\bar E_{+}\otimes e^{\bar\Sigma}+1\otimes \bar E_{+}
\\\nonumber
&& \\\nonumber
\Delta_{5}(\bar H)
&= & \bar H\otimes e^{-\bar\Sigma}+1\otimes \bar H\,  %-(-1)^k \frac{\chi^\prime}{\chi_{k+1}}\, E_{k+} e^{-\Sigma_k}\otimes\Sigma_{k+1} e^{-\Sigma_k}
%\nonumber \\[8pt]&&
- \frac{\rho}{4}\,\left(\H\otimes \bar E_{+} e^{-\bar\Sigma}-\, \bar E_{+} e^{-\bar\Sigma}\otimes \H e^{-\bar\Sigma}\right)
\\\nonumber
&& \\\nonumber
 \Delta_{5}(\bar E_{-})&=&\bar E_{-}\otimes e^{-\bar\Sigma}+1\otimes \bar E_{-}
+2\bar\chi\bar H\otimes \bar H e^{-\bar\Sigma}
 +\bar\chi\bar H(\bar H-1)\otimes\bar\Lambda+
\nonumber \\[8pt]&&
\,-\frac{\rho}{2}\,\left( \bar H e^{-\bar\Sigma}\otimes\H e^{-\bar\Sigma} - \bar H\H \otimes\bar\Lambda
-\H \otimes \bar H e^{-\bar\Sigma}\right)
\nonumber \\[8pt]&&
\,-\frac{\rho}{2}\, \left(\,
\bar\Lambda e^{\bar\Sigma}\otimes \bar H\H e^{-\bar\Sigma}\,+
\bar H\bar\Lambda e^{\bar\Sigma}\otimes\H \bar\Lambda\,
 \right)\nonumber \\[8pt]&&
 \,-\,\frac{\rho}{4}\,\left(
 %2H_k\Lambda_ke^{\Sigma_k} +%-2\,\Lambda_k \Sigma_{k+1} e^{\Sigma_k}
 \left(1-e^{-2\bar\Sigma}\right)\otimes\H \bar\Lambda\, +\H \otimes\bar\Lambda\,-\bar\Lambda\otimes
\H e^{-\bar\Sigma}\right) \nonumber\\[8pt]
&&
{1\over\bar\chi}\left(\frac{\rho}{4}\right)^2\,\left(\bar{\Lambda}^2 e^{2\bar\Sigma}\otimes\H^2 \bar\Lambda
+\bar\Lambda\otimes\H^2 e^{-\bar\Sigma}
+\H^2\otimes\bar\Lambda
\right)
 \nonumber\\[8pt]&&
- {2\over\bar\chi}\left(\frac{\rho}{4} \right)^2\,\bar\Lambda \H e^{\bar\Sigma}\otimes\H\bar\Lambda\,
\end{eqnarray}
The antipodes do not depend on the Abelian twist and look as follows:
\begin{eqnarray}\label{blte22b }
S_{5_{}}^{}(q_{}^{\pm \H_{}})&=&q_{}^{\mp \H_{}}~,\qquad
S_{5_{}}^{}(\E_{\pm})\;=\;-q_{}^{\pm1}\E_{\pm}~,\nonumber\\[8pt]
S_{5}(\bar E_{+})&=&-\bar E_{+}\,e^{-\bar\Sigma}  ,\qquad\qquad  S_{5}(\bar H)=-\bar He^{\bar\Sigma}\\[8pt]
  S_{5}(\bar E_{-})&=&-\bar E_{-}\,e^{\bar\Sigma} +\bar\chi \bar{H}^2 e^{\bar\Sigma}(e^{\bar\Sigma}+1)
  -\bar{\chi}^2 \bar H \bar E_{+}e^{\bar\Sigma} .\nonumber%(e^{\Sigma_k}-1)\nonumber
  \end{eqnarray}
%The antipods for $\bar H,\, \bar E,\,\bar F$ are obtained by complex conjugation.
%The quantum  tree (complex) parameters $R$-matrix $R_2\equiv R_{2^{}}^{}(\chi,\bar\chi,\varsigma)$ takes, in this case, the form
%\begin{equation}\label{qr2}
%R_{2^{}}^{}=\exp{(\frac{-\varsigma}{\chi\bar\chi}\Sigma\wedge\bar\Sigma)}\exp{(\Sigma\otimes H)} \exp{(-H\otimes \Sigma)} \exp{(\bar\Sigma \otimes \bar H)}  \exp{(-\bar H\otimes \bar\Sigma)} \exp{(\frac{-\varsigma}{\chi\bar\chi}\Sigma\wedge\bar\Sigma)}.
%\end{equation}
Quantum universal $R$-matrix generated from $r_V$ takes the following form
\begin{equation}\label{R5}
R_5(\gamma,\bar\chi,\rho)= \tilde q^{\bar\Sigma\wedge \H} R_{3'0}(\gamma)F^\tau_{J1}(\bar\chi)F^{-1 }_{J1}(\bar\chi) \tilde q^{\bar\Sigma\wedge \H}.
\end{equation}
Three real quantizations we describe in the Table \ref{tab:rr5} below.

\begin{center}
% {\bf Table A. All real Lie bialgebra structures for $\mathfrak{o}(4;\mathbb{C})$ }\\[1ex]
\begin{table}[h]
\begin{tabular}{|c|c|c|c|c|}
\hline
&&&&\\[-8pt]
$\mathfrak{o'}^{\star}(4)$ &  $\gamma,\rho\in\mathbb{R}$ ; $\bar\chi\in\imath\mathbb{R}$  &$(q^\H)^{\dag}=q^\H, \E_{\pm}^{\dag}=\E_{\mp}$
& $\bar H^{\star}= -\bar H , \bar E_{\pm}^{\star}=-\bar E_{\pm}$& R\\
\hline
&&&\\[-8pt]
$\mathfrak{o}'(2,2)$ &  $\gamma,\rho\in\mathbb{R}$ ; $\bar\chi\in\imath\mathbb{R}$ &$(q^\H)^{\#}=q^\H, \E_{\pm}^{\#}=-\E_{\mp}$
& $\bar H^{\star}= -\bar H , \bar E_{\pm}^{\star}=-\bar E_{\pm}$& R\\
\hline
&&&\\[-8pt]
$\mathfrak{o}''(2,2)$ &    $\gamma,\rho, \bar\chi\in\imath\mathbb{R}$   &
$(q^\H)^{\star}=q^\H, \E_{\pm}^{\star}=-\E_{\pm}$
& $\bar{H}^{\star}=-\bar H , \bar E_{\pm}^{\star}=-\bar E_{\pm}$& A\\
\hline
 \end{tabular}
\caption{Real quantizations of $r_{V}(\gamma, \rho, \bar\chi)=
\gamma\,E_{+}\wedge E_{-} +\bar{\chi}\,\bar{E}_{+}\wedge\bar{H}+\rho H\wedge\bar{E}_{+}$}
	\label{tab:rr5}
\end{table}
\end{center}

\section{Concluding remarks and outlook}

In this paper we presented the complete set of Hopf-algebraic quantum deformations generated by classical $r$-matrices for $\mathfrak{o}(4;\mathbb{C})$ and its real forms given in \cite{BLTnov15,BLT17}.  The explicit formulae describing algebraic and coalgebraic sectors are provided as well as there are given the universal $R$-matrices  which permits the tensoring of quantum modules (representations of quantum-deformed  Hopf algebras). We recall that the universal $R$-matrices describe the braided structure of quantum-covariant tensor products of modules \cite{Pod92,Maj95} what has been used in quantum-covariant NC field theory \cite{FW07,JLMW}. For quantum twist deformations of enveloping Lie algebras 
$U(\mathfrak{g})$ ($\mathfrak{g}=\mathfrak{o}(4;\mathbb{C}), \mathfrak{o}(4-k,k)$ ($k=0,1,2$), 
and $\mathfrak{o}^*(4)=\mathfrak{0}(2;\mathbb{H})$); for   $\mathfrak{o}(4;\mathbb{C})$ 
%obtained by twisting  (i.e. $\gamma=\bar\gamma=0$ in (\ref{r1})-(\ref{r6})), 
(see Sect.5.1, 5.2) the algebra of quantum modules, describing e.g. NC quantum fields, can be represented by the functions of classical (commutative) fields with twist-dependent nonlocal star product multiplication rule \cite{Bloh03,Kul05}.

%
%The $D=4$ rotational symmetries with their quantum deformations can be applied at least in three different ways:
%
%i) as quantum deformed rotational symmetries of various $D=4$ quantum spacetimes, with arbitrary metric signatures or quaternionic structure.

Basic role plays in relativistic physics $D=4$ Minkowski space $\mathbb{R}^{3,1}$, with signature $(+,+,+,-)$, and its Lorentz rotations $\mathfrak{o}(3,1)$. In order to describe the Lie algebra generating relativistic group of motion, one adds  four generators $P_\mu\in \mathbb{T}^{3,1}$ of translations, i.e. one extends Lorentz algebra  $\mathfrak{o}(3,1)$ to $D=4$ 
 Poincar\'{e} algebra $\mathfrak{o}(3,1)\ltimes \mathbb{T}^{3,1}$.
It is known that only two out of four quantum deformations of $\mathfrak{o}(3,1)$ can be extended to quantum deformations of $D=4$ Poincar\'{e} algebra (see \cite{Zak94,Zak97,Tol07,BLT08}). The studies providing the complete list of possible quantum deformations of inhomogeneous  $D=4$ Euclidean  $\mathfrak{o}(4)\ltimes \mathbb{T}^{4}$ algebra and of inhomogeneous  $D=4$ Kleinian $\mathfrak{o}(2,2)\ltimes \mathbb{T}^{2,2}$ algebra has not been presented
\footnote{For partial results in $D=4$ Euclidean case see e.g. \cite{BLT15}}.%-\cite{BLT15}, \cite{Paolo}.}.
We add that inhomogeneous extension of quaternionic real form $\mathfrak{o}^*(4)\equiv \mathfrak{o}(2;\mathbb{H})$
of $\mathfrak{o}(4; \mathbb{C})$ contains four complex or two quaternionic translations and its applications to the description of physical symmetries are, according to our knowledge, not known.

The real forms of considered  quantum groups describe the quantum symmetries of $D=3$ compact Euclidean ($S^3$), 
de Sitter ($dS_3$) or anti-de-Sitter ($AdS_3$) spacetimes, with finite nonvanishing constant curvature %(cosmological constant) 
and curved $D=3$ Euclidean, dS$_3$ or AdS$_3$ translations.
In $D=4$ rotation algebras $\mathfrak{o}(3,1)$  ($\mathfrak{o}(2,2)$), the dS (AdS) radius $\mathcal{R}$ is introduced by suitable rescaling of the generators in the coset $\frac{\mathfrak{o}(4-k,k)}{\mathfrak{o}(3-k,k)}$ ($k=1,2$), 
with $\Lambda=\mathcal{R}^{-1}$ which can be treated as a deformation parameter. The quantum deformations of $ \mathfrak{o}(4-k,k)$ ($k=1,2$) have been extensively studied as describing the NC geometry of $2+1$-dimensional QG with cosmological constant $\Lambda\neq 0$ 
\cite{Witten88,Ball14}. The classical action of $D=3$ gravity can be introduced geometrically as gauge theory 
described by $D=3$ Chern-Simons (CS) model. Following Fock-Rosly construction \cite{Fock99,Alek95}, in such framework we describe gravitational degrees of freedom as parameterizing the Poisson-Lie group manifold. If we search for quantum deformations of Fock-Rosly construction, it appears that only classical $r$-matrices obtained from Drinfeld double (DD) structures \cite{Dr2} are allowed \cite{BS09,BS10}. The DD structures and corresponding classical $r$-matrices for $\mathfrak{o}(3,1)$ and $\mathfrak{o}(2,2)$ algebras were recently constructed and classified \cite{Ball14}. We see that such quantum deformation which are well adjusted to the description of quantum-deformed $D=3$ gravity are generated by a subclass of 
classical $r$-matrices, listed in \cite{BLTnov15,BLT17} and quantized in this paper.

%iii) One can consider complex group $\mathfrak{o}(4;\mathbb{C})$ and its respective real forms as the target space of principal $\sigma$-models. We add that two-dimensional $\sigma$-models are used for the description  of (super)string action 
%\cite{Tsey98}. The class of deformed string models were subsequently described by so-called Yang-Baxter $\sigma$-models 
%\cite{Klim02}-\cite{BKLSY15}, with kinetic bilinear part of the action  modified by the suitable insertion of classical $r$-matrices.
%The presented list of classical $r$-matrices can be used %for all real forms of $\mathfrak{o}(4;\mathbb{C})$, there are mentioned several applications 
%to the description of quantum deformations of $D=10$ superstrings with particular choices of the target spaces, containing the sectors with $\mathfrak{o}(4)\equiv S^3\times S^3, \mathfrak{o}(3,1)\equiv dS_3$  
%and $\mathfrak{o}(2,2)\equiv AdS_3$ symmetries.

The next  step in our program is  to construct the complete list of classical $r$-matrices for the $D=4$ complex inhomogeneous Euclidean algebra $\mathcal{E}(4;\mathbb{C}):=\mathfrak{io}(4;\mathbb{C}):=\mathfrak{o}(4;\mathbb{C})\ltimes\mathbf{T}(4;\mathbb{C})$ (orthogonal rotations together with translations) and for its real forms,
in particular $\mathfrak{o}(4-k,k)\ltimes\mathbf{T}(4-k,k;\mathbb{R})$  ($k=0,1,2$).
Until present time the most complete results were obtained for $\mathfrak{o}(3,1)\ltimes\mathbf{T}(3,1)$ by Zakrzewski 
\cite{Zak97}, who provided almost complete list of 21 different, not related by Poincar\'{e} automorphism real $D=4$ Poincar\'e $r$-matrices  (see also \cite{Tol07,BP14}).
It should be noticed that the complete classifications of r-matrices for both inhomogenous $D=3$  Poincar\'{e} and  $D=3$
Euclidean algebras have been given by Stachura \cite{Stach98}. %One-parameter $\kappa$-deformation of $D=3$  Poincar\'{e}
%symmetry has been recently introduced in \cite{Beisert17}.

Recently in \cite{BLMT12,BLMT11} the present authors  complexified Zakrzewski results and then imposed $D=4$ Euclidean reality constraints. It appeared that 8 out of 21 complexified  Zakrzewski $r$-matrices are consistent with the Euclidean conjugation in $\mathfrak{o}(4;\mathbb{C})$ (see (\ref{rf1})). %(see (\ref{crr2})). 
It can be shown, however, that the complexified Zakrzewski $r$-matrices do not describe all $r$-matrices for $\mathcal{E}(4;\mathbb{C})$\footnote{In particular one can easily argue observing that the list of the real $r$-matrices for $\mathfrak{o}(2,2)\ltimes\mathbf{T}(2,2)$ is longer then the Zakrzewski list (see \cite{Zak97}) for  $D=4$ Poincar\'e algebra.}. Using the constructive method analogous to the one proposed in this paper we intend to describe the complete classification of classical $r$-matrices for $D=4$ complex inhomogeneous Lie algebra 
$\mathcal{E}(4;\mathbb{C})$ and for its all real forms.

We add that in \cite{BLMT12,BLMT11} we considered also the $N=1$ superextension of Poincar\'e and Euclidean classical $r$-matrices. Recently we derived in analogous way as well new class of $N=2$ Poincar\'e and Euclidean supersymmetric $r$-matrices (see \cite{BLT15}). We hope that our constructive method of providing the complete list of classical $r$-matrices for the complex $\mathcal{E}(4;\mathbb{C})$ case 
%and then describing its all real forms 
can be applied as well to $N$-extended Euclidean superalgebras $\mathcal{E}(4|N;\mathbb{C})$ for $N=1,2,4$ 
and further classify and quantize 
the supersymmetric $r$-matrices for the corresponding real forms.%, in particular for physically important $N=4$ case.

\subsection*{Acknowledgments}
%The authors would like to thank P.~Aschieri,  S.~van Tongeren and K.~Yoshida for valuable comments.
This work has been supported by Polish National Science Center (NCN), project 2014/13/B/ST2/04043 (A.B. and J.L.) and by COST (European Cooperation in Science and Technology) Action MP1405 QSPACE. V.N.T. was supported by RFBR grant No.14-01-00474-a.
\appendix
\section{All $\mathfrak{o}(3)$ and $\mathfrak{o}(2,1)$ Lie bialgebras}
\renewcommand{\theequation}{A.\arabic{equation}}
\setcounter{equation}{0}
Classification of r-matrices is the same task as classification of coboundary Lie bialgebras up to isomorphisms (an isomorphism which preserves the structure constants is called an automorphism). In geometric terms they can be seen as orbits of an action of the Lie algebra automorphism group in the space of skew-symmetric solution of mCYBE. For simple algebras all bialgebra structures are coboundary due to the Whitehead lemma.

Let us consider, for completeness as well as for pedagogical reason, geometric classification scheme for classical r-matrices of simple 3-dimensional real rotational Lie algebras (for purely algebraic approach see \cite{LT17}). Up to an isomorphism there are only two non-isomorphic real simple Lie algebras: compact $\mathfrak{o}(3)$ and non-compact $\mathfrak{o}(2,1)$,
%$$\mathfrak{o}(3)\cong\mathfrak{su}(2)\quad \mbox{and}\quad \mathfrak{o}(2,1)\cong\mathfrak{o}(1,2)\cong\mathfrak{sl}(2;\mathbb{R})\cong\mathfrak{su}(1,1)$$
both are real form of $\mathfrak{o}(3;\mathbb{C})$.

Consider firstly the compact $\mathfrak{o}(3)$ case with the canonical vectorial  basis ($I_k^\dag=-I_k$ cf. (\ref{jl2}) -- (\ref{jl3b}))
 \begin{equation}\label{a1}
	 [I_1, I_2]=I_3, \quad [I_1, I_3]= - I_2, \quad [I_2, I_3]=I_1
 \end{equation}
We notice that any element $r(a,b,c)= a I_2\wedge I_3+ b  I_3\wedge I_1+c I_1\wedge I_2 \in \mathfrak{o}(3)\wedge\mathfrak{o}(3)$ is a classical r-matrix since it satisfies 
\begin{equation}\label{a2}
	[[r(a,b,c), r(a,b,c)]]= (a^2+b^2+c^2)\Omega
\end{equation}
where $\Omega=I_1\wedge I_2\wedge I_3\in \mathfrak{o}(3)\wedge\mathfrak{o}(3)\wedge\mathfrak{o}(3)$ is a unique up to the constant invariant element.
 %\medskip\\

The non-isomorphic Lie bialgebra structures for $\mathfrak{o}(3)$ case can be identify with orbits of the automorphism group in the space of free parameters $(a,b,c)\in \mathbb{R}^3$ with the Euclidean metric. The group of automorphisms contain $SO(3)$ subgroup. Moreover, bivector and vector representations are equivalent in dimension 3. Due to this property we look only for
$SO(3)$- orbits in $\mathbb{R}^3$. These are the 2-dimensional spheres represented by a radius $\xi>0$ or by the vector $\xi(1,0,0)$. Thus as a result of final classification
one gets the following family of non-trivial $\mathfrak{o}(3)$ r-matrices (the trivial $r=0$ $r$-matrix corresponds to singular one-point orbit at $(0,0,0)$)
\begin{equation}\label{a3}
	r_\xi= \xi I_1\wedge I_2
\end{equation}
Notice that the values of the real parameter $\xi>0$ are effective and lead to nonequivalent Lie bialgebra structures.

Similar analysis applied to the non-compact real form $\mathfrak{o}(2,1)$  provides qualitatively different results.
Arbitrary $\mathfrak{o}(2,1)$ r-matrix satisfies the following YB equation (a,b,c real)
\begin{equation}\label{a4}
	[[r(a,b,c), r(a,b,c)]]= (a^2-b^2 +c^2) J_1\wedge J_2\wedge J_3
\end{equation}
where  $r(a,b,c)= a J_2\wedge J_3+ b  J_3\wedge J_1+c J_1\wedge J_2 \in \mathfrak{o}(2,1)\wedge\mathfrak{o}(2,1)$ is written in
the canonical $\mathfrak{o}(2,1)$ basis. We choose noncompact vectorial generators $J_1, J_2, J_3$
 following our choice of $\mathfrak{o}(2,1)$ reality conditions  $J_k^\star=-J_k$
\begin{equation}\label{a5}
	 [J_1, J_2]=J_3, \quad [J_1, J_3]=  J_2, \quad [J_2, J_3]=  J_1
 \end{equation}
The automorphisms group $SO(2,1)$ of the Lie algebra $\mathfrak{o}(2,1)$  acts in three-dimensional Minkowski space
$\mathbb{R}^{2,1}$. There are three types of non-trivial orbits in  $\mathbb{R}^{2,1}=\{(a,b,c): a,b,c\in\mathbb{R}\}$, characterizing three
 independent $\mathfrak{o}(2,1)$  r-matrices.

1. single light-cone orbit represented by the light-like vector $\xi(1,1,0)$  which provides the solution
of homogeneous classical YB equation (CYBE)

2. one-parameter family of space-like orbits represented by  space-like vectors\\ $\xi (1,0,0), \xi\neq 0$
 (solution of modified CYBE)

3. one-parameter families of time-like orbits represented by time-like vectors\\ $\xi (0,1,0),
\xi\neq 0 $
 (solution of modified CYBE)

Three  canonical $\mathfrak{o}(2,1)$ r-matrices corresponding to three types of orbits  take the form
\begin{equation}\label{a6}
r^{0}_\xi= \xi (J_1\wedge J_3+   J_1\wedge J_2);
\quad r^{-}_\xi= \xi  J_3\wedge J_2;
\quad r^{+}_\xi=  \xi J_1\wedge J_3
\end{equation}
where in the first case
 one gets the same deformation for any
 value of the parameter $\xi\neq 0$,  while in the remaining two cases different  values of $\xi$ lead  to different
 Lie bialgebras. In this setting we get one reality condition and three different types of $r$-matrices representing nonequivalent bialgebra structures.

%As it has been shown in \cite{LT17} all these cases can be, by suitable transformations (being real Lie algebra isomorphisms), 
%written down in the Cartan-Weyl generators in such a way that the standard deformation has always the standard  $r_{st}= E_+\wedge E_-$ form. For example, defining $H=J_1, E_{\pm}=J_3\pm J_2$, one gets $r_J=H\wedge E_+=J_1\wedge (J_3+J_2)$ 
%(cf. (\ref{rm6-4})) and $r_{st}= E_+\wedge E_- = 2 J_2\wedge J_3$ together with $\mathfrak{sl}(2)$-reality conditions 
%$H^\star=-H, (E_\pm)^\star=-E_\pm$ (cf. (\ref{rm6-3})).
%Taking instead $H=\imath J_2, E_{\pm}=J_1\mp \imath J_3$ we will get $\mathfrak{su}(1,1)$ reality condition and $E_+\wedge E_-=\imath J_1\wedge J_3$ (cf. (\ref{rm6-2})). Similarly,  we can obtain (\ref{rm6-1}) from (\ref{a3}) \cite{LT17}.

%Finally, we can use the (real) Lie algebra isomorphisms $\mathfrak{o}(2, 1)\cong \mathfrak{sl}(2)\cong\mathfrak{su}(1,1)$ to transform these results to another basis (cf. \cite{LT17} for details).

\section{$q$-exponent and $q$-Hadamard formula}
\renewcommand{\theequation}{B.\arabic{equation}}
\setcounter{equation}{0}
 
Our aim here is to introduce some formulas (mainly concerning  a $q$-deformed Hadamard lemma), 
%\footnote{Hadamard lemma is sometimes called Baker-Campbell-Hausdorff formula.}, 
which were main tools for calculations  presented in Sect. 5.4.

Let $A$ and $B$ be two arbitrary elements of some quantum algebra and let $\exp_{q}(A)$
be a formal $q$-exponential %(\ref{ct16}) 
%where we use the standard definition of $q$-exponential $\exp_{q_{}^{-2}}$ (cf. Appendix B)
\begin{eqnarray}\label{q-exp}
\exp_{q}(A)\!\!&:=\!\!&\sum_{n\geq0}\,\frac{A^n}{(n)_{q}^{}!}~,
\quad\;(n)_{q}^{}!:=(1)_{q}^{}(2)_{q}^{}\cdots
(n)_{q}^{},\quad(n)_{q}^{}=\frac{1-q^n}{1-q}~.
\end{eqnarray}
of the element $A$. As the $q$-exponential $\exp_{q^{-1}}(-A)$ is inverse to $\exp_{q}(A)$, i.e. $\bigl(\exp_{q}(A)\bigr)^{-1}=
\exp_{q^{-1}}(-A)$ thus the $q$-analog of  Hadamard formula can be obtained as follows (see \cite{KhTo1})
\begin{eqnarray}\label{A1}
\begin{array}{rcl}
\exp_{q}(A)\,B\bigl(\exp_{q}(A)\bigr)^{-1}\!\!&=\!\!&\exp_{q}(A)\,B\exp_{q^{-1}}(-A)\;
\equiv\;\bigl(\mathop{\rm Ad}\exp_{q}(A)\bigr)(B)\;=\;
\\[10pt]
\!\!&=\!\!&\displaystyle\Bigl(\sum_{n\geq 0}\frac{1}{(n)_{q}!}(\mathop{{\rm
ad}_q}A)^n\Bigr)(B)\;=\;\bigl(\exp_{q}(\mathop{{\rm ad}_q}A)\bigr)(B)~,
\end{array}
\end{eqnarray}
where the $q$-adjoint action is defined by means of $q$-brackets ($[C,\,D]_{q'}\;\equiv\;CD-q'DC$~):
\begin{eqnarray}\label{B2}
\begin{array}{rcl}
(\mathop{{\rm ad}_q}A)^0(B)\!\!&\equiv\!\!&B~,\quad(\mathop{{\rm ad}_q}(A))^1(B)\;\equiv
\;[A,\,B]~,\quad(\mathop{{\rm ad}_q}(A))^2(B)\;\equiv\;[A,\,[A,\,B]]_{q}~,
\\[10pt]
(\mathop{{\rm ad}_q}(A))^3(B)\!\!&\equiv\!\!&[A,\,[A,\,[A,\,B]]_{q}]_{q^2}~,\ldots,
(\mathop{{\rm ad}_q}(A))^{n+1}(B)\;=\;[A,\,(\mathop{{\rm ad}_q}(A))^{n}(B)]_{q^n}~.
\end{array}
\end{eqnarray}
Consider the spacial case ($q'\neq q$ in general)
\begin{eqnarray}\label{A3}
[A,\,B]_{q'}\!\!&=\!\!&0~,
\end{eqnarray}
one gets
\begin{eqnarray}\label{A4a}
\begin{array}{rcl}
(\mathop{{\rm
ad}_q}(A))^{n+1}(B)\!\!&=\!\!&\displaystyle(1-q'^{-1}q^{n})A\bigl(\mathop{{\rm
ad}_q}(A)\bigr)^{n}(B)\;=\;\prod_{k=0}^{n}(1-q'^{-1}q^{k})A^{n}B
\\[10pt]
\!\!&=\!\!&(q'^{-1};q)_{n}\,A^{n}B~.
\end{array}
\end{eqnarray}
using  the standard notation $(a;q)_n$ from the theory of basic hypergeometric
series (see e.g. \cite{Klimyk,GaRa}). Substituting (\ref{A4a}) in  (\ref{A1}) we obtain
\begin{eqnarray} \label{A5}
\begin{array} {rcl}
\exp_{q}(A)\,B\bigl(\exp_{q}(A)\bigr)^{-1}\!\!&=\!\!&\displaystyle\Bigl(
\sum_{n=0}^{\infty} \frac{(q'^{-1};q)_{n}}{(q;q)_n}\,(1-q)^nA^{n}\Bigr)B\;=\;
\\[10pt]
\!\!&=\!\!&\displaystyle{}_1\phi_{0}(q'^{-1};-;q;(1-q)A)\,B\;=\;\frac{\bigl(q'^{-1}
(1-q)A;q\Bigr)_{\infty}}{((1-q)A;q)_{\infty}}\,B~,
\end{array}
\end{eqnarray}
as  the result of the $q$-binomial theorem (see \cite{Klimyk,GaRa}). 
%Thus we obtained that {\it $q$-Hadamard formula (\ref{A1}) provided (\ref{A3})
%expresses in terms of the $q$-binomial formula, i.e.}
%\begin{eqnarray}\label{A6}
%\exp_{q}(A)\,B(\exp_{q}(A))^{-1}\!\!&=\!\!&{}_1\phi_{0}\bigl(q'^{-1};-;q;(1-q)
%A\bigr)\,B\;=\;\frac{\bigl(q'^{-1}(1-q)A;q\bigr)_{\infty}}{\bigl((1-q)A;q\bigr)_{\infty}}\,B~.
%\end{eqnarray}
In the particular case $q'=q^{n}$, $n=0,1,2,\ldots,$ the formula (\ref{A1}) reads
\begin{eqnarray}\label{A7}
\begin{array}{rcl}
\exp_{q}(A)\,B(\exp_{q}(A))^{-1}\!\!&=\!\!&\bigl(q^{-n}(1-q)A;q\bigr)_{n}\,B
\\[10pt]
\!\!&=\!\!&q^{-n(n+1)/2}\bigl((q-1)A\bigr)^{n}\,\bigl(q/(1-q)A;q\bigr)_{n}\,B~.
\end{array}
\end{eqnarray}
In the case $q'=q^{-n}$, $n=1,2,\ldots,$ for (\ref{A1}) we have
\begin{eqnarray}\label{A8}
\exp_{q}(A)\,B(\exp_{q}(A))^{-1}\!\!&=\!\!&\bigl((1-q)A;q\bigr)_{n}^{-1}\,B~.
\end{eqnarray}
%Simple utilization of the formulas (\ref{A7}) and (\ref{A8}) gives rise to the formulae
%(\ref{BD8})--(\ref{BD14}) and (\ref{BD17})--(\ref{BD21}).
%%More specifically one needs
%In some cases this formula can be written in the finite (compact) form:
%\begin{itemize}
%\item {\em If $[A, B]_q=0$ then}
%\begin{eqnarray}\label{A3}
%\exp_{q}(A)\,B\bigl(\exp_{q}(A)\bigr)^{-1}\!\!&=\!\!&\tilde{X}B\;=\;BX~.
%\end{eqnarray}
%\item {\em If  $[A, B]_{q^{-1}}=0$ then}
%\begin{eqnarray}\label{A4}
%\exp_{q}(A)\,B\bigl(\exp_{q}(A)\bigr)^{-1}\!\!&=\!\!&X^{-1}B\;=\;B\tilde{X}^{-1}~.
%\end{eqnarray}
%{\em where $X=1+(q-1)A$,  $\tilde X=1-(q^{-1}-1)A$}~.
%\end{itemize}
Adopting to the situation in Sect. 5.4  one has to substitute $A\rightarrow\mathbb{A}=\varsigma
\E_{+}q_{}^{\H_{}+\bar \H_{}}\otimes q_{}^{\H_{}+\bar \H_{}}\bar \E_{+}$ or $A=\varsigma \E_{+}\bar \E_{+}$
and $n=1,2$ (cf. (\ref{BD15}) or (\ref{BD17})).

\end{document}